\documentclass[longauth]{aa}

\usepackage{graphicx}
\usepackage{txfonts}
\usepackage{comment}
\usepackage{float}
\usepackage{textcomp}
\usepackage{xcolor}

\begin{document}

   \title{GRAVITY K-band spectroscopy of HD~206893~B\thanks{Based on observations made with ESO telescopes at Paranal Observatory under program IDs 1103.B-0626 and 1104.C-0651.}}

\subtitle{Brown dwarf or exoplanet}

\author{J.~Kammerer\inst{\ref{esog},\ref{australie},\ref{stsci}}
 \and S.~Lacour\inst{\ref{lesia},\ref{esog}}
 \and T.~Stolker\inst{\ref{leiden}}
 \and P.~Molli\`ere\inst{\ref{mpia}}
 \and D.~K.~Sing \inst{\ref{jhuastro},\ref{jhueps}}
 \and E.~Nasedkin\inst{\ref{mpia}}
 \and P.~Kervella\inst{\ref{lesia}}
 \and J.~J.~Wang\inst{\ref{caltech}}
 \and K.~Ward-Duong\inst{\ref{stsci}}
 \and M.~Nowak\inst{\ref{cam}}
 \and R.~Abuter\inst{\ref{esog}}
 \and A.~Amorim\inst{\ref{lisboa},\ref{centra}}
 \and R.~Asensio-Torres\inst{\ref{mpia}}
 \and M.~Baub\"ock\inst{\ref{mpe}}
 \and M.~Benisty\inst{\ref{ipag}}
 \and J.-P.~Berger\inst{\ref{ipag}}
 \and H.~Beust\inst{\ref{ipag}}
 \and S.~Blunt\inst{\ref{caltech}}
 \and A.~Boccaletti\inst{\ref{lesia}}
 \and A.~Bohn\inst{\ref{leiden}}
 \and M.-L.~Bolzer\inst{\ref{mpe}}
 \and M.~Bonnefoy\inst{\ref{ipag}}
 \and H.~Bonnet\inst{\ref{esog}}
 \and W.~Brandner\inst{\ref{mpia}}
 \and F.~Cantalloube\inst{\ref{mpia}}
 \and P.~Caselli \inst{\ref{mpe}}
 \and B.~Charnay\inst{\ref{lesia}}
 \and G.~Chauvin\inst{\ref{ipag}}
 \and E.~Choquet\inst{\ref{lam}}
 \and V.~Christiaens\inst{\ref{monash}}
 \and Y.~Cl\'enet\inst{\ref{lesia}}
 \and V.~Coud\'e~du~Foresto\inst{\ref{lesia}}
 \and A.~Cridland\inst{\ref{leiden}}
 \and R.~Dembet\inst{\ref{esog},\ref{lesia}}
 \and J.~Dexter\inst{\ref{colorado}}
 \and P.~T.~de~Zeeuw\inst{\ref{leiden},\ref{mpe}}
 \and A.~Drescher\inst{\ref{mpe}}
 \and G.~Duvert\inst{\ref{ipag}}
 \and A.~Eckart\inst{\ref{cologne},\ref{bonn}}
 \and F.~Eisenhauer\inst{\ref{mpe}}
 \and F.~Gao\inst{\ref{hambourg}}
 \and P.~Garcia\inst{\ref{centra},\ref{porto}}
 \and R.~Garcia~Lopez\inst{\ref{dublin},\ref{mpia}}
 \and E.~Gendron\inst{\ref{lesia}}
 \and R.~Genzel\inst{\ref{mpe}}
 \and S.~Gillessen\inst{\ref{mpe}}
 \and J.~Girard\inst{\ref{stsci}}
 \and X.~Haubois\inst{\ref{esoc}}
 \and G.~Hei\ss el\inst{\ref{lesia}}
 \and T.~Henning\inst{\ref{mpia}}
 \and S.~Hinkley\inst{\ref{exeter}}
 \and S.~Hippler\inst{\ref{mpia}}
 \and M.~Horrobin\inst{\ref{cologne}}
 \and M.~Houll\'e\inst{\ref{lam}}
 \and Z.~Hubert\inst{\ref{ipag},\ref{lesia}}
 \and L.~Jocou\inst{\ref{ipag}}
 \and M.~Keppler\inst{\ref{mpia}}
 \and L.~Kreidberg\inst{\ref{mpia}}
 \and A.-M.~Lagrange\inst{\ref{ipag},\ref{lesia}}
 \and V.~Lapeyr\`ere\inst{\ref{lesia}}
 \and J.-B.~Le~Bouquin\inst{\ref{ipag}}
 \and P.~L\'ena\inst{\ref{lesia}}
 \and D.~Lutz\inst{\ref{mpe}}
 \and A.-L.~Maire\inst{\ref{liege},\ref{mpia}}
 \and A.~M\'erand\inst{\ref{esog}}
 \and J.~D.~Monnier\inst{\ref{umich}}
 \and D.~Mouillet\inst{\ref{ipag}}
 \and A.~M\"uller\inst{\ref{mpia}}
 \and T.~Ott\inst{\ref{mpe}}
 \and G.~P.~P.~L.~Otten\inst{\ref{lam},\ref{ntu}}
 \and C.~Paladini\inst{\ref{esoc}}
 \and T.~Paumard\inst{\ref{lesia}}
 \and K.~Perraut\inst{\ref{ipag}}
 \and G.~Perrin\inst{\ref{lesia}}
 \and O.~Pfuhl\inst{\ref{esog}}
 \and L.~Pueyo\inst{\ref{stsci}}
 \and J.~Rameau\inst{\ref{ipag}}
 \and L.~Rodet\inst{\ref{cornell}}
 \and G.~Rousset\inst{\ref{lesia}}
 \and Z.~Rustamkulov \inst{\ref{jhueps}}
 \and J.~Shangguan\inst{\ref{mpe}}
 \and T.~Shimizu \inst{\ref{mpe}}
 \and J.~Stadler\inst{\ref{mpe}}
 \and O.~Straub\inst{\ref{mpe}}
 \and C.~Straubmeier\inst{\ref{cologne}}
 \and E.~Sturm\inst{\ref{mpe}}
 \and L.~J.~Tacconi\inst{\ref{mpe}}
 \and E.F.~van~Dishoeck\inst{\ref{leiden},\ref{mpe}}
 \and A.~Vigan\inst{\ref{lam}}
 \and F.~Vincent\inst{\ref{lesia}}
 \and S.~D.~von~Fellenberg\inst{\ref{mpe}}
 \and F.~Widmann\inst{\ref{mpe}}
 \and E.~Wieprecht\inst{\ref{mpe}}
 \and E.~Wiezorrek\inst{\ref{mpe}}
 \and J.~Woillez\inst{\ref{esog}}
 \and S.~Yazici\inst{\ref{mpe}}
 \and  the GRAVITY Collaboration}
\institute{
   European Southern Observatory, Karl-Schwarzschild-Stra\ss e 2, 85748 Garching, Germany
\label{esog}      \and
   Research School of Astronomy \& Astrophysics, Australian National University, Canberra, ACT 2611, Australia
\label{australie}      \and
   LESIA, Observatoire de Paris, PSL, CNRS, Sorbonne Universit\'e, Universit\'e de Paris, 5 place Janssen, 92195 Meudon, France
\label{lesia}      \and
   Leiden Observatory, Leiden University, P.O. Box 9513, 2300 RA Leiden, The Netherlands
\label{leiden}      \and
   Max Planck Institute for Astronomy, K\"onigstuhl 17, 69117 Heidelberg, Germany
\label{mpia}      \and
   Five College Astronomy Department, Amherst College, Amherst, MA 01002, USA
\label{amherst}      \and
   Department of Astronomy, California Institute of Technology, Pasadena, CA 91125, USA
\label{caltech}      \and
   Institute of Astronomy, University of Cambridge, Madingley Road, Cambridge CB3 0HA, United Kingdom
\label{cam}      \and
   Universidade de Lisboa - Faculdade de Ci\^encias, Campo Grande, 1749-016 Lisboa, Portugal
\label{lisboa}      \and
   CENTRA - Centro de Astrof\' isica e Gravita\c c\~ao, IST, Universidade de Lisboa, 1049-001 Lisboa, Portugal
\label{centra}      \and
   Max Planck Institute for extraterrestrial Physics, Giessenbachstra\ss e~1, 85748 Garching, Germany
\label{mpe}      \and
   Universit\'e Grenoble Alpes, CNRS, IPAG, 38000 Grenoble, France
\label{ipag}      \and
   Aix Marseille Univ, CNRS, CNES, LAM, Marseille, France
\label{lam}      \and
   School of Physics and Astronomy, Monash University, Clayton, VIC 3800, Melbourne, Australia
\label{monash}      \and
  JILA and Department of Astrophysical and Planetary Sciences, University of Colorado, Boulder, CO 80309, USA
\label{colorado}      \and
   1. Institute of Physics, University of Cologne, Z\"ulpicher Stra\ss e 77, 50937 Cologne, Germany
\label{cologne}      \and
   Max Planck Institute for Radio Astronomy, Auf dem H\"ugel 69, 53121 Bonn, Germany
\label{bonn}      \and
   Hamburger Sternwarte, Universit\"at Hamburg, Gojenbergsweg 112, 21029 Hamburg, Germany
\label{hambourg}      \and
   Universidade do Porto, Faculdade de Engenharia, Rua Dr. Roberto Frias, 4200-465 Porto, Portugal
\label{porto}      \and
   School of Physics, University College Dublin, Belfield, Dublin 4, Ireland
\label{dublin}      \and
   Astronomy Department, University of Michigan, Ann Arbor, MI 48109 USA
\label{umich}      \and
   Space Telescope Science Institute, Baltimore, MD 21218, USA
\label{stsci}      \and
   European Southern Observatory, Casilla 19001, Santiago 19, Chile
\label{esoc}      \and
   University of Exeter, Physics Building, Stocker Road, Exeter EX4 4QL, United Kingdom
\label{exeter}      \and
   STAR Institute/Universit\'e de Li\`ege, Belgium
\label{liege}      \and
   Center for Astrophysics and Planetary Science, Department of Astronomy, Cornell University, Ithaca, NY 14853, USA
\label{cornell}    \and
Department of Physics and Astronomy, Johns Hopkins University, Baltimore, MD, USA
\label{jhuastro}  \and
Department of Earth \& Planetary Sciences, Johns Hopkins University, Baltimore, MD, USA
\label{jhueps} \and
Academia Sinica, Institute of Astronomy and Astrophysics, 11F Astronomy-Mathematics Building, NTU/AS campus, No. 1, Section 4, Roosevelt Rd., Taipei 10617, Taiwan
\label{ntu}.
}

   \date{Received September 15, 1996; accepted March 16, 1997}


  \abstract
   {Near-infrared interferometry has become a powerful tool for studying the orbital and atmospheric parameters of substellar companions.}
   {We aim to reveal the nature of the reddest known substellar companion HD~206893~B by studying its near-infrared colors and spectral morphology and by investigating its orbital motion.}
   {We fit atmospheric models for giant planets and brown dwarfs and perform spectral retrievals with \texttt{petitRADTRANS} and \texttt{ATMO} on the observed GRAVITY, SPHERE, and GPI spectra of HD~206893~B. To recover its unusual spectral features, first and foremost its extremely red near-infrared color, we include additional extinction by high-altitude dust clouds made of enstatite grains in the atmospheric model fits. However, forsterite, corundum, and iron grains predict similar extinction curves for the grain sizes considered here. We also infer the orbital parameters of HD~206893~B by combining the $\sim 100~\text{\textmu as}$ precision astrometry from GRAVITY with data from the literature and constrain the mass and position of HD~206893~C based on the \emph{Gaia} proper motion anomaly of the system.}
   {The extremely red color and the very shallow $1.4~\text{\textmu m}$ water absorption feature of HD~206893~B can be fit well with the adapted atmospheric models and spectral retrievals. By comparison with AMES-Cond evolutionary tracks, we find that only some atmospheric models predict physically plausible objects. Altogether, our analysis suggests an age of $\sim 3$--$300~\text{Myr}$ and a mass of $\sim 5$--$30~\text{M}_\text{Jup}$ for HD~206893~B, which is consistent with previous estimates but extends the parameter space to younger and lower-mass objects. The GRAVITY astrometry points to an eccentric orbit ($e = 0.29^{+0.06}_{-0.11}$) with a mutual inclination of $< 34.4~\text{deg}$ with respect to the debris disk of the system.}
   {While HD~206893~B could in principle be a planetary-mass companion, this possibility hinges on the unknown influence of the inner companion on the mass estimate of $10^{+5}_{-4}~\text{M}_\text{Jup}$ from radial velocity and \emph{Gaia} as well as a relatively small but significant Argus moving group membership probability of $\sim 61\%$. However, we find that if the mass of HD~206893~B is $< 30~\text{M}_\text{Jup}$, then the inner companion HD~206893~C should have a mass between $\sim 8$--$15~\text{M}_\text{Jup}$. Finally, further spectroscopic or photometric observations at higher signal-to-noise and longer wavelengths are required to learn more about the composition and dust cloud properties of HD~206893~B.}


   \maketitle
%

\section{Introduction}
\label{sec:introduction}

\begin{figure*}
\centering
\includegraphics[trim=0cm 6cm 0cm 4cm, clip, width=\textwidth]{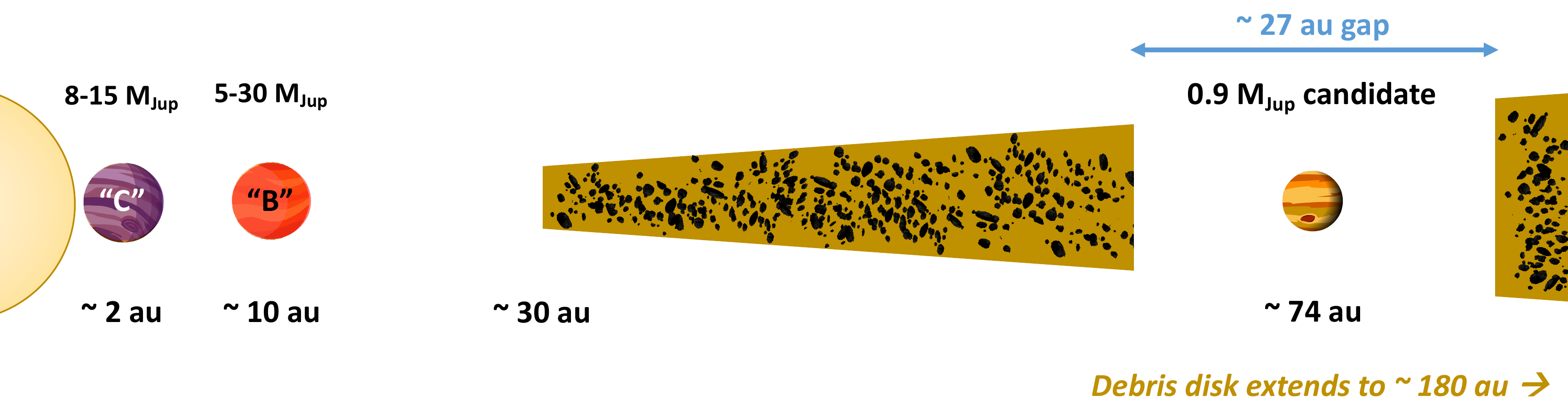}
\caption{Schematic of the HD~206893 system with the two inner companions ``B'' and ``C,'' the debris disk with its $\sim 27~\text{au}$ wide gap, and the planetary-mass companion candidate at $\sim 74~\text{au}$ that could be responsible for clearing this gap. The quoted mass estimates for HD~206893~B and~C are explained in Section~\ref{sec:mass} of this paper. Distances and sizes are not to scale.}
\label{fig:schematic}
\end{figure*}

While the number of known exoplanets has grown to over 4000 in the last decade\footnote{\url{http://exoplanet.eu/}}, the sample of directly imaged substellar companions remains small \citep[e.g.,][]{bowler2016}. These objects are prime targets for a direct study of their atmospheric properties and composition through imaging and low-resolution spectroscopy \citep[e.g.,][]{biller2018}, though. Together with evolutionary tracks and atmospheric models, this enables their effective temperature, radius, surface gravity, age, and mass to be inferred and conclusions to be drawn on their formation history and subsequent evolution \citep[e.g.,][]{bowler2016,biller2018}. Moreover, astrometric measurements from direct imaging enable deriving the orbital parameters of substellar companions and studying the dynamical interactions with their environment \citep[e.g.,][]{kley2012}.

Recently, \citet{gravity2019} and \citet{gravity2020} used long-baseline interferometry to perform medium-resolution ($R \sim 500$) K-band spectroscopy of exoplanets. They were able to demonstrate astrometric measurements with a precision $\sim 10$ times better than what had previously been possible and constrain the atmospheric C/O ratio of the gas giant $\beta$~Pic~b. This, together with a chemical abundance framework of its protoplanetary disk based on \citet{oeberg2011}, enabled them to infer a formation by warm-start core accretion \citep{pollack1996} between the water and carbon-dioxide icelines. Furthermore, \citet{nowak2020} were able to directly detect $\beta$~Pic~c, another gas giant in the same system previously discovered with the radial velocity technique \citep{lagrange2019}. They showed that the precise mass estimate from the radial velocity data together with the K-band spectrum also favors a formation by warm-start core accretion \citep[e.g.,][]{mordasini2013} for $\beta$~Pic~c. In this regard, near-infrared interferometry significantly advances the field of planet formation and evolution by enabling direct observations of exoplanets discovered with the radial velocity technique for the first time ever.

Another system for which both direct observations and radial velocity data are available is HD~206893. This debris disk system, at a distance of $40.8~\text{pc}$ \citep{gaia2018}, hosts a directly detected substellar companion at a separation of $\sim 10~\text{au}$, HD~206893~B \citep{milli2017}. Astronomers are puzzled by the nature of this companion due to its unusually red near-infrared color. \citet{delorme2017} found that an additional K-band extinction of $\sim 0.5~\text{mag}$ is required to match the spectrum of HD~206893~B with those of other dusty, low-gravity, or young brown dwarfs. Furthermore, \citet{delorme2017} and \citet{ward-duong2020} showed that its extremely red color together with its very shallow $1.4~\text{\textmu m}$ water absorption feature are challenging to fit with currently available atmospheric models without an additional extra-photospheric source of extinction.

While \citet{delorme2017} argue for HD~206893~B being an extremely dusty $15$--$30~\text{M}_\text{Jup}$ L-dwarf, consistent with their age estimate of $50$--$700~\text{Myr}$ for the host star, \citet{ward-duong2020} note that its H- and K-band spectra suggest a lower surface gravity and younger object. Together with the high infrared excess of the disk and a possible Argus moving group membership \citep[membership probability $\sim 61\%$;][]{ward-duong2020}, there is a consistent scenario for HD~206893~B being a young ($< 50~\text{Myr}$) gas giant planet. This scenario is also supported by the dynamical mass estimate of $10^{+5}_{-4}~\text{M}_\text{Jup}$ from \citet{grandjean2019} based on radial velocity data and the \emph{Gaia} proper motion anomaly of the system. Their analysis also points to a second, closer-in companion at a separation of $\sim 1.4$--$2.6~\text{au}$ (HD~206893~C). The picture is further complicated by a gap in the debris disk of the system, which was recently discovered using the Atacama Large Millimeter/Submillimeter Array and could be carved by a third, Jupiter-mass companion at $\sim 74~\text{au}$ \citep{marino2020}. A schematic of the HD~206893 system is shown in Figure~\ref{fig:schematic} for illustrative purposes.

Here, we present Very Large Telescope Interferometer (VLTI)/GRAVITY K-band spectroscopy of HD~206893~B. From these GRAVITY data, we extract the astrometry of HD~206893~B with a precision of $\sim 100~\text{\textmu as}$ and a medium-resolution ($R \sim 500$) K-band contrast spectrum, which we convert to a spectrum of HD~206893~B with a model spectrum of its host star (cf. Section~\ref{sec:observations_and_data_reduction}). From the astrometry, we improve the constraints on the orbital parameters of HD~206893~B (cf. Section~\ref{sec:astrometric_analysis}) and on the mass and position of HD~206893~C, also using the \emph{Gaia} proper motion anomaly of the system. Furthermore, we perform atmospheric model fitting, with and without additional extra-photospheric extinction by high-altitude dust clouds made of enstatite grains (cf. Section~\ref{sec:atmospheric_model_fitting}). A similar study investigating the dynamical mass and the circumplanetary accretion flux of the PDS~70~b and~c protoplanets using GRAVITY data has recently been published by \citet{wang2021}. Moreover, we perform spectral retrievals for HD~206893~B (cf. Section~\ref{sec:spectral_retrieval}) and check for consistency between our best fit atmospheric parameters and evolutionary tracks (cf. Section~\ref{sec:evolutionary_tracks}). Finally, we discuss our findings in the context of previous works on this system (cf. Section~\ref{sec:discussion}).


\section{Observations and data reduction}
\label{sec:observations_and_data_reduction}

\begin{table*}
\caption{Observing log. NEXP, NDIT, and DIT denote the number of exposures, the number of detector integrations per exposure, and the detector integration time, respectively, and $\tau_0$ denotes the atmospheric coherence time.}
\label{tab:log}
\centering
\begin{tabular}{ccccccccc}
\hline\hline
Date & \multicolumn{2}{c}{UT time} & \multicolumn{2}{c}{NEXP/NDIT/DIT} & Airmass & $\tau_0$ & Seeing \\
& Start & End & HD~206893~B & HD~206893~A & & & \\
\hline
2019-07-17 & 08:52:56 & 09:56:06 & 5/12/60~s & 6/64/1~s & 1.17--1.54 & 1.5--2.2~ms & $1.11$--$1.71^{\prime\prime}$ \\
2019-08-13 & 03:21:16 & 04:21:09 & 5/12/60~s & 7/64/1~s & 1.03--1.12 & 3.2--4.9~ms & $0.80$--$1.00^{\prime\prime}$ \\
\hline
\end{tabular}
\end{table*}

\begin{table}
\caption{Relative astrometry of HD~206893~B.}
\label{tab:astrometry}
\centering
\begin{tabular}{cccccc}
\hline\hline
MJD & $\Delta\text{RA}$ & $\Delta\text{Dec}$ & $\sigma_{\Delta\text{RA}}$ & $\sigma_{\Delta\text{Dec}}$ & $\rho$ \\
(days) & (mas) & (mas) & (mas) & (mas) & -- \\
\hline
58681.396 & 130.73 & 198.10 & 0.04 & 0.06 & -0.58 \\
58708.165 & 127.03 & 199.27 & 0.09 & 0.13 & -0.88 \\
\hline
\multicolumn{6}{l}{\textbf{Notes.} \parbox[t]{7 cm}{The covariance matrix can be obtained using $\sigma_{\Delta\text{RA}}^2$ and $\sigma_{\Delta\text{Dec}}^2$ on the diagonal and $\rho\sigma_{\Delta\text{RA}}\sigma_{\Delta\text{Dec}}$ off-diagonal, where $\rho$ is the correlation coefficient.}}
\end{tabular}
\end{table}

\begin{table}
\caption{Stellar parameters and 2MASS and and WISE photometry of HD~206893~A from the literature.}
\label{tab:hd206893a_parameters}
\centering
\begin{tabular}{cccc}
\hline\hline
Parameter & Unit & Value & Source \\
\hline
$T_\text{eff}$ & K & $6500 \pm 100$ & D17 \\
$\log g$ & -- & $4.45 \pm 0.15$ & D17 \\
$[\text{Fe/H}]$ & dex & $0.04 \pm 0.02$ & D17 \\
$R$ & $\text{R}_\odot$ & $1.26 \pm 0.02$ & D17 \\
$\pi$ & mas & $24.51 \pm 0.06$ & G18 \\
\hline
$\text{J}_\text{2MASS}$ & mag & $5.869 \pm 0.023$ & S06 \\
$\text{H}_\text{2MASS}$ & mag & $5.687 \pm 0.034$ & S06 \\
$\text{Ks}_\text{2MASS}$ & mag & $5.593 \pm 0.021$ & S06 \\
$\text{W1}_\text{WISE}$ & mag & $5.573 \pm 0.176$ & W10 \\
$\text{W2}_\text{WISE}$ & mag & $5.452 \pm 0.052$ & W10 \\
$\text{W3}_\text{WISE}$ & mag & $5.629 \pm 0.015$ & W10 \\
$\text{W4}_\text{WISE}$ & mag & $5.481 \pm 0.043$ & W10 \\
\hline
\multicolumn{4}{l}{\textbf{Notes.} \parbox[t]{5.5 cm}{D17 = \citet{delorme2017}, G18 = \citet{gaia2018}, S06 = \citet{skrutskie2006}, W10 = \citet{wright2010}.}}
\end{tabular}
\end{table}

\begin{figure}
\centering
\includegraphics[width=\columnwidth]{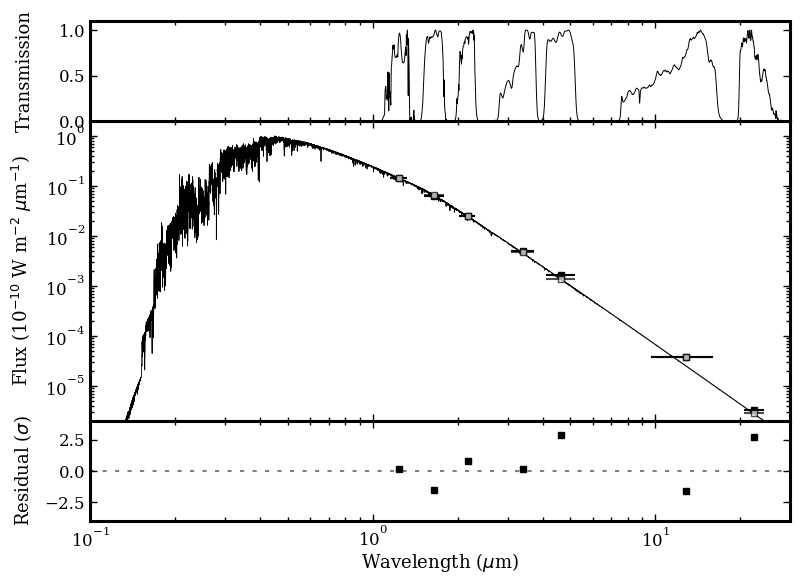}
\caption{BT-NextGen model spectrum of HD~206893~A (black line), scaled to fit the shown 2MASS and WISE photometry (gray data points). Synthetic photometry based on the model spectrum is also shown with black data points. The top panel shows the transmission curves corresponding to each photometric data point and the bottom panel shows the residuals between the synthetic and the observed photometry.}
\label{fig:hd206893a_spectrum}
\end{figure}

As part of the ExoGRAVITY large program \citep{lacour2020}, we obtained two epochs of GRAVITY \citep{gravity2017} medium-resolution ($R \sim 500$) K-band spectra of HD~206893~A and~B using the Unit Telescopes (UTs) at the VLTI. The observing log is presented in Table~\ref{tab:log}. The atmospheric conditions varied between average (atmospheric coherence time $\tau_0 = 5~\text{ms}$) and below average ($\tau_0 \approx 1.5~\text{ms}$). Both epochs were obtained as bad weather backup targets for the large active galactic nucleus program (PI E.~Sturm) based on a time exchange agreement.

From the raw GRAVITY data, we extracted the coherent flux following the standard recipe of the official ESO data reduction pipeline \citep{lapeyrere2014}. During this first step, the pipeline computes the coherent flux observed on the host star and the companion. However, the coherent flux observed on the faint companion is still contaminated by the halo of the bright host star. This contamination was removed using a Python package developed by our team\footnote{Python package available on GitHub upon request.}. The individual steps of this package are outlined in Appendix~A of \citet{gravity2020} and its output is the decontaminated ratio of the coherent flux between the companion and the host star.

The astrometry for each epoch of data was obtained from the phase of the ratio of the coherent fluxes and is presented in Table~\ref{tab:astrometry}. The uncertainties were estimated from the scatter of the astrometric values obtained independently for each individual exposure. The typical precision is $\sim 100~\text{\textmu as}$, which is much worse than the theoretical limit of $16.5~\text{\textmu as}$ determined by \citet{lacour2014}. This can be attributed to the low and high frequency phase errors present in our data and introduced by instrumental aberrations \citep{gravity2021}. Furthermore, due to the asymmetry of the uv-plane, we used the correlation coefficient $\rho$ to properly describe the confidence intervals and the correlations between the right ascension and the declination offset.

Finally, a spectrum of the companion for each epoch of data was obtained from the amplitude of the ratio of the coherent fluxes. The host star is essentially unresolved in our observations such that we did not need to correct this ratio for the visibility of the host star. The ratio of the coherent fluxes was multiplied by a model spectrum of the host star, which was obtained by interpolating the BT-NextGen stellar model grid \citep{allard2012} for the stellar parameters presented in Table~\ref{tab:hd206893a_parameters}. This yields a spectrum of HD~206893~B for each epoch of data. The model spectrum of the host star was scaled to match the stellar photometry presented in Table~\ref{tab:hd206893a_parameters}, yielding a flux calibrated radius of $1.36~\text{R}_\odot$. The 2MASS and WISE photometry are sufficient to constrain the stellar spectrum over the GRAVITY wavelength range (cf. Figure~\ref{fig:hd206893a_spectrum}). Here, the stellar spectrum is essentially smooth and the weak spectral lines are negligible for our analysis.


\section{Orbit analysis}
\label{sec:orbit_analysis}

\subsection{Astrometric analysis}
\label{sec:astrometric_analysis}

\begin{table*}
\caption{Orbital parameters inferred for HD~206893~B. The posterior states the 68\% confidence interval around the median. $\mathcal{N}(\mu,\sigma)$ denotes a normal distribution with mean $\mu$ and standard deviation $\sigma$. For comparison, the orbital parameters inferred by other works are shown as well.}
\label{tab:hd206893b_orbit}
\centering
\begin{tabular}{ccccccccc}
\hline\hline
Parameter & Description & Prior & This work$^{(a)}$ & G19$^{(a)}$ & W21$^{(a)}$ & This work$^{(b)}$ & M20$^{(b)}$ \\
\hline
$a$ (au) & Semimajor axis & LogUniform(1, 100) & $9.28^{+1.77}_{-0.93}$ & $9.74^{+1.46}_{-1.41}$ & $10.4^{+1.8}_{-1.7}$ & $11.37^{+1.09}_{-0.75}$ & $11.35^{+1.13}_{-0.77}$ \\
$e$ & Eccentricity & Uniform(0, 1) & $0.29^{+0.06}_{-0.11}$ & $0.25^{+0.13}_{-0.17}$ & $0.23^{+0.13}_{-0.16}$ & $0.13^{+0.05}_{-0.03}$ & $0.14^{+0.05}_{-0.04}$ \\
$i$ (deg) & Inclination & Sine(0, 180)$^{(a)}$ & $154^{+12}_{-9}$ & $146^{+13}_{-7}$ & $146^{+14}_{-7}$ & -- & -- \\
& & $\mathcal{N}(140, 3)^{(b)}$ & -- & -- & -- & $142^{+2}_{-3}$ & $140^{+2}_{-2}$ \\
$\omega$ (deg) & Argument of per. & Uniform(0, 360) & $123^{+89}_{-44}$ & $181^{+223}_{-135}$ & $173^{+79}_{-109}$ & $71^{+36}_{-25}$ & $74^{+36}_{-29}$ \\
$\Omega$ (deg) & Lon. of asc. node & Uniform(0, 180)$^{(a)}$ & $55^{+49}_{-28}$ & $267^{+44}_{-29}$ & $77^{+34}_{-33}$ & -- & -- \\
& & $\mathcal{N}(61, 4)^{(b)}$ & -- & -- & -- & $60^{+3}_{-4}$ & $62^{+4}_{-4}$ \\
$\tau$ & Time of per. since $\tau_\text{ref}^{(c)}$ & Uniform(0, 1) & $0.58^{+0.10}_{-0.20}$ & -- & -- & $0.32^{+0.12}_{-0.09}$ & -- \\
$\pi$ (mas) & Parallax & $\mathcal{N}(24.51, 0.06)^\text{(d)}$ & $24.50^{+0.06}_{-0.06}$ & $24.51^{+0.06}_{-0.06}$ & $24.51^{+0.06}_{-0.06}$ & $24.51^{+0.06}_{-0.06}$ & $24.51^{+0.06}_{-0.06}$ \\
$M_\text{tot}$ ($\text{M}_\odot$) & Total mass & $\mathcal{N}(1.32, 0.02)^\text{(e)}$ & $1.32^{+0.02}_{-0.02}$ & $1.32^{+0.02}_{-0.02}$ & $1.32^{+0.02}_{-0.02}$ & $1.32^{+0.02}_{-0.02}$ & $1.32^{+0.02}_{-0.02}$ \\
$P$ (yr) & Orbital period & -- & $24.59^{+7.38}_{-3.60}$ & $26.45^{+6.18}_{-5.54}$ & $29.1^{+8.1}_{-6.7}$ & $33.36^{+4.96}_{-3.30}$ & $33.28^{+5.09}_{-3.33}$ \\
\hline
\multicolumn{8}{l}{\textbf{Notes.} \parbox[t]{17 cm}{Scenario 1 is constrained by the data only and scenario 2 has an additional constraint on coplanarity with the debris disk. (a) Scenario 1, (b) Scenario 2, (c) Measured in fractions of orbital period with $\tau_\text{ref} = 55800$ MJD, (d) \citealt{gaia2018}, (e) \citealt{delorme2017}, G19 = \citealt{grandjean2019}, W21 = \citealt{ward-duong2020}, M20 = \citealt{marino2020}.}}
\end{tabular}
\end{table*}

\begin{figure*}
\centering
\includegraphics[width=0.85\textwidth]{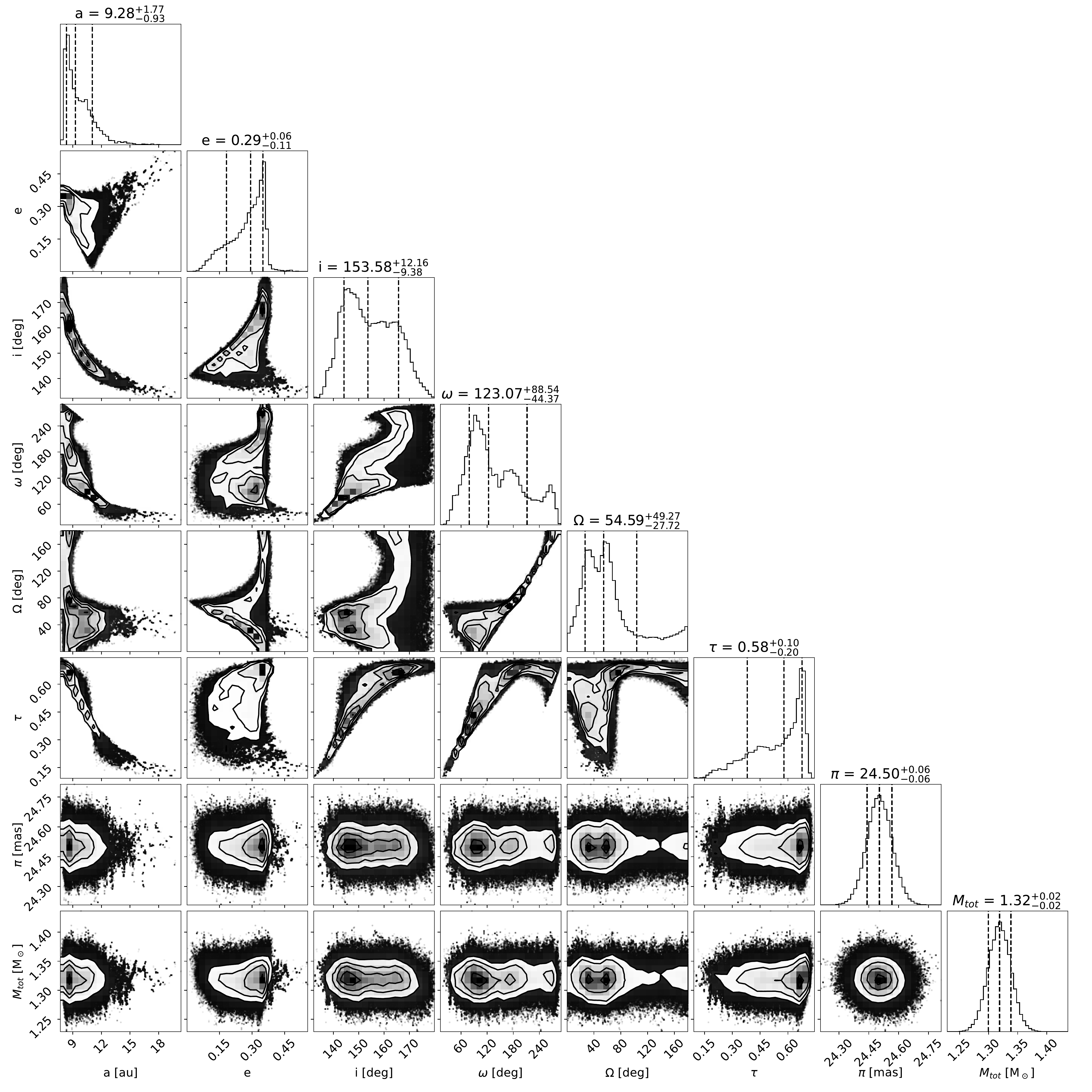}
\includegraphics[width=0.85\textwidth]{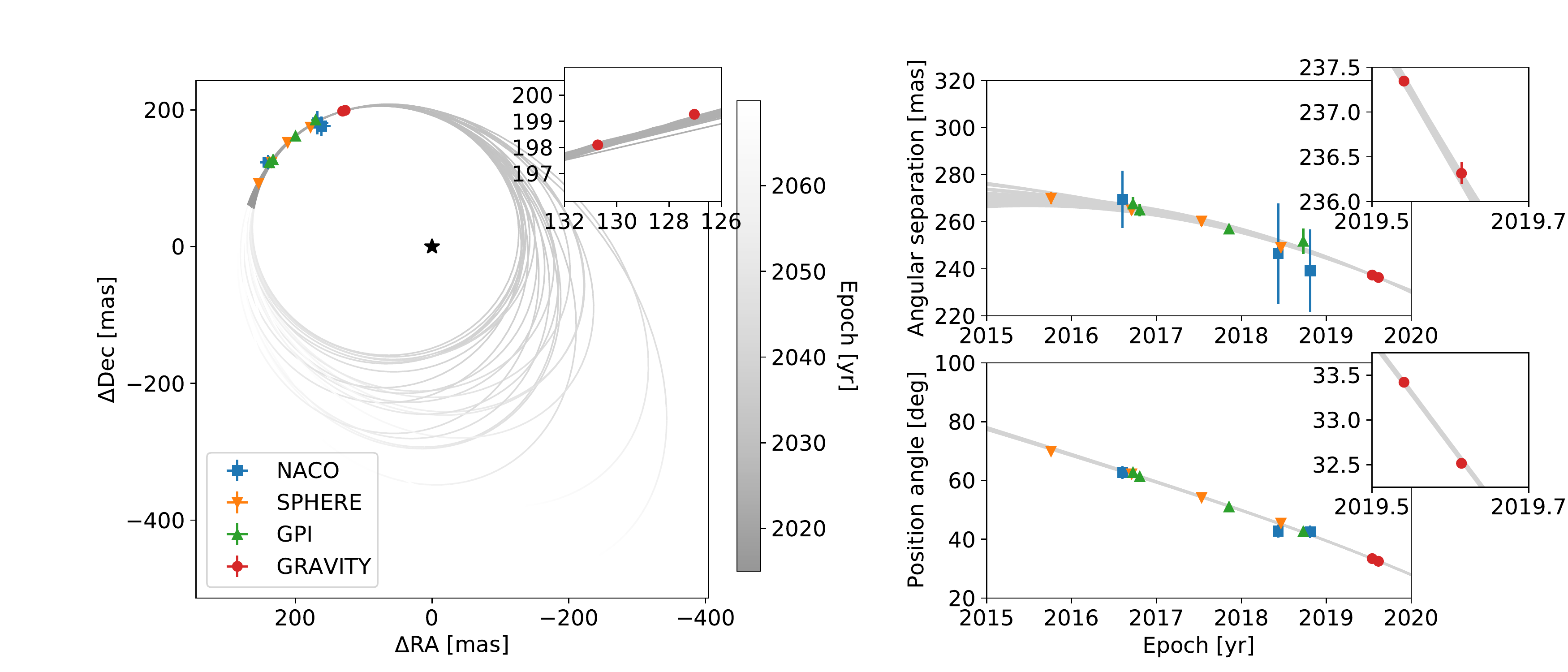}
\caption{Posterior distribution of the orbital parameters (top) and orbital solutions together with the NACO, SPHERE, GPI, and GRAVITY astrometry (bottom) of HD~206893~B. In the top panel, the values state the 68\% confidence intervals around the median. In the bottom panel, the black star highlights the position of HD~206893~A and all error bars show the 1--$\sigma$ confidence intervals.}
\label{fig:orbit_fitting_free}
\end{figure*}

From the interferometric observations with GRAVITY we obtain two new astrometric data points with an unprecedented precision of $\sim 100~\text{\textmu as}$ for HD~206893~B (cf. Table~\ref{tab:astrometry}). Together with astrometric data from the literature (cf. Table~\ref{tab:hd206893b_astrometry}), we estimate the orbital parameters of HD~206893~B with \texttt{orbitize!}\footnote{\url{https://github.com/sblunt/orbitize}} \citep{blunt2020}, which infers the posterior distribution of the orbital parameters through Markov Chain Monte Carlo (MCMC) sampling with \texttt{ptemcee}\footnote{\url{https://github.com/willvousden/ptemcee}} \citep{foreman-mackey2013,vousden2016}. We initialize the sampler with 20 temperatures, 500 walkers, and 50000 steps per walker. By visual inspection of the walker chains, we assess convergence and reject the first 40000 steps before computing the posterior distribution from each walker at the lowest temperature. We note that we use the priors presented in Table~\ref{tab:hd206893b_orbit} for the orbital parameters. These priors are chosen very conservatively in order to not constrain the posterior. The posterior distribution of the orbital parameters and the inferred orbital solutions together with the NACO, SPHERE, GPI, and GRAVITY astrometry are shown in Figure~\ref{fig:orbit_fitting_free} and the posterior values are quoted in Table~\ref{tab:hd206893b_orbit} (scenario 1). We restrict our orbit analysis to direct imaging astrometry of HD~206893~B because \citet{grandjean2019} noted that the radial velocities and the \emph{Gaia} astrometry can only be explained when including an inner companion (HD~206893~C) in the fit. However, including this inner companion could introduce biases in our fit since its orbital parameters and dynamical mass remain poorly constrained.

In general, the GRAVITY astrometry is consistent with those from NACO, SPHERE, and GPI and our orbital solutions are consistent with those from \citet{grandjean2019} and \citet{ward-duong2020}. Both of them obtained a double-peaked semimajor axis distribution and an anticorrelation between semimajor axis $a$ and eccentricity $e$. However, adding the GRAVITY astrometry disfavors small eccentricities $e \sim 0$ and removes the second peak in the semimajor axis distribution. Instead, a smaller semimajor axis of $9.28^{+1.77}_{-0.93}~\text{au}$ and a higher eccentricity of $0.29^{+0.06}_{-0.11}$ are preferred. Nevertheless, the inclination $i$ is similar to that obtained by \citet{grandjean2019} and \citet{ward-duong2020} with a maximum likelihood value of $\sim 145~\text{deg}$. The mutual inclination $i_\text{m}$ between the debris disk of the system reported by \citet{milli2017} and \citet{marino2020} and HD~206893~B is
\begin{equation}
    i_\text{m} = \arccos\left(\cos(i)\cos(i_\text{d})+\sin(i)\sin(i_\text{d})\cos(\Omega-\Omega_\text{d})\right),
\end{equation}
where $i_\text{d} = 140 \pm 3~\text{deg}$ is the inclination of the debris disk and $\Omega_\text{d} = 61 \pm 4~\text{deg}$ is the longitude of the ascending node of the debris disk \citep{marino2020}. We compute the mutual inclination and its uncertainty by drawing samples from the posterior distribution of the orbit fit and a normal distribution for $i_\text{d}$ and $\Omega_\text{d}$. We find $i_\text{m} = 20.8^{+13.6}_{-11.2}~\text{deg}$, which means that the debris disk and the companion are roughly aligned. We note that we use priors between $0$ and $180~\text{deg}$ for the longitude of the ascending node $\Omega$ to enforce the debris disk and the companion orbiting in the same direction. However, the direction of rotation of the debris disk is unconstrained and there is a $180~\text{deg}$ ambiguity in the mutual inclination $i_\text{m}$ between the debris disk and the companion \citep{heintz1978}. Therefore, they might as well orbit in opposite directions.

Interestingly, \citet{marino2020} found that if they enforce coplanarity between the debris disk and the companion, this would lead to the other one of the degenerate solutions being preferred, namely a larger semimajor axis of $\sim 11.4~\text{au}$ and a smaller eccentricity of $\sim 0.14$. To verify their findings, we ran another fit with Gaussian priors of $140 \pm 3~\text{deg}$ for the inclination $i$ and $61 \pm 4~\text{deg}$ for the longitude of the ascending node $\Omega$. The results are shown in Figure~\ref{fig:orbit_fitting_disk} and Table~\ref{tab:hd206893b_orbit} (scenario 2) and confirm the findings of \citet{marino2020}, even when adding the GRAVITY astrometry. Therefore, depending on whether coplanarity with the debris disk is assumed or not, degenerate orbital solutions are obtained for HD~206893~B.

\subsection{\emph{Gaia} proper motion anomaly}
\label{sec:gaia_proper_motion_anomaly}

\begin{figure*}
\centering
\includegraphics[width=\textwidth]{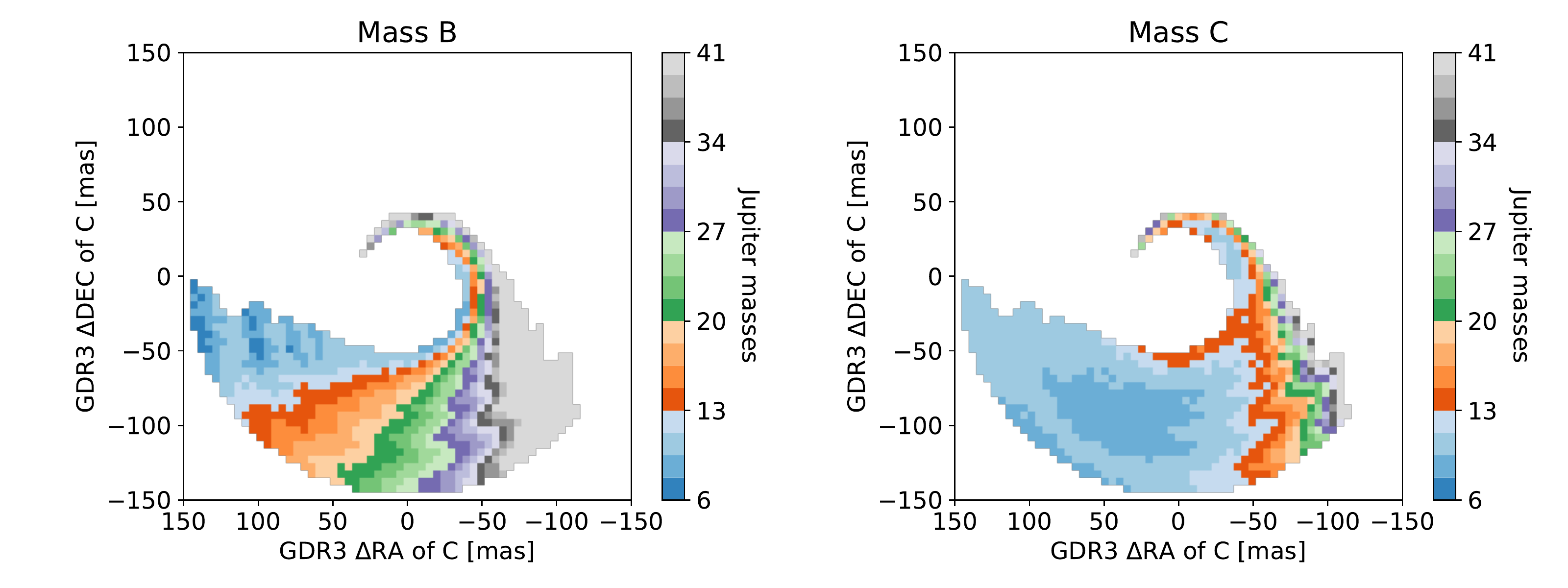}
\caption{Masses of HD~206893~B (left) and~C (right) as a function of the on-sky position of HD~206893~C (at the reference epoch of \emph{Gaia} EDR3, 2016.0) derived from the proper motion anomaly measured between the \emph{Gaia} EDR3 proper motion and the Hipparcos--\emph{Gaia} EDR3 long-term proper motion. Physically implausible regions due to negative masses or dynamical instability have been ignored.}
\label{fig:gaia_mass}
\end{figure*}

An independent constraint on the mass of HD~206893~B can be obtained from the proper motion anomaly of its host star (HD~206893~A) measured by \emph{Gaia} \citep{kervella2020}. While the proper motion measured by \emph{Gaia} \citep{gaia2018} traces the photocenter of the system, the long-term proper motion derived from the difference between the \emph{Gaia} and \emph{Hipparcos} \citep{vanleeuwen2007} positions traces the barycenter of the system, once corrected for the influence of the companion at the time of the \emph{Hipparcos} and \emph{Gaia} measurements. The difference between the proper motion measured by \emph{Gaia} and the long-term proper motion is the proper motion anomaly that must be caused by one or multiple faint companions.

\citet{grandjean2019} have already found that the proper motion anomaly measured between the \emph{Gaia} DR2 and the \emph{Hipparcos} data cannot be explained by HD~206893~B alone and suggested the presence of another closer-in companion of $\sim 15~\text{M}_\text{Jup}$ (HD~206893~C). Here, we consider the proper motion anomaly measured between the \emph{Gaia} EDR3 and the \emph{Hipparcos} data ($\text{PMa}_\text{RA} = -0.102 \pm 0.037$ and $\text{PMa}_\text{DEC} = -0.612 \pm 0.028$) to obtain independent constraints on the masses of HD~206893~B and~C and the on-sky position of HD~206893~C. Since the orbit of HD~206893~B is known from direct observations (cf. Section~\ref{sec:astrometric_analysis}) we can compute the proper motion anomaly it causes on its host star. Consistently with \citet{grandjean2019}, we find that the proper motion anomaly measured by \emph{Gaia} EDR3 cannot be explained by HD~206893~B alone because a proper motion component tangential to the one caused by HD~206893~B is necessary to fit the data. Hence, we assume another companion (HD~206893~C) on a similarly oriented orbit (same inclination and longitude of the ascending node as HD~206893~B), but with zero eccentricity and smaller semimajor axis. We compute the proper motion anomaly it causes on its host star as a function of its on-sky position and its mass, marginalizing over 100 randomly drawn samples of the posterior distribution of the orbit of HD~206893~B and the uncertainties in the proper motion anomaly measured by \emph{Gaia} EDR3.

Figure~\ref{fig:gaia_mass} shows the predicted masses for HD~206893~B and~C. Regions where either of the companions would have negative mass, where the orbital period of HD~206893~C would be smaller than half of the \emph{Gaia} EDR3 measurement timespan, or where the apparent separation of HD~206893~C would be more than $150~\text{mas}$ are ignored. Our orbital fits for HD~206893~B suggest a minimum orbital separation of $\sim 150~\text{mas}$ and we assume that the orbits of the two companions cannot cross for dynamical stability reasons. For most of the on-sky positions, the mass of HD~206893~C should be between $\sim 8$--$15~\text{M}_\text{Jup}$, a result that is largely consistent with the prediction of \citet{grandjean2019}. Furthermore, depending on the mass of HD~206893~B, the on-sky position of HD~206893~C can be strongly constrained.


\section{Spectral analysis}
\label{sec:spectral_analysis}

\begin{figure*}
\centering
\includegraphics[width=\textwidth]{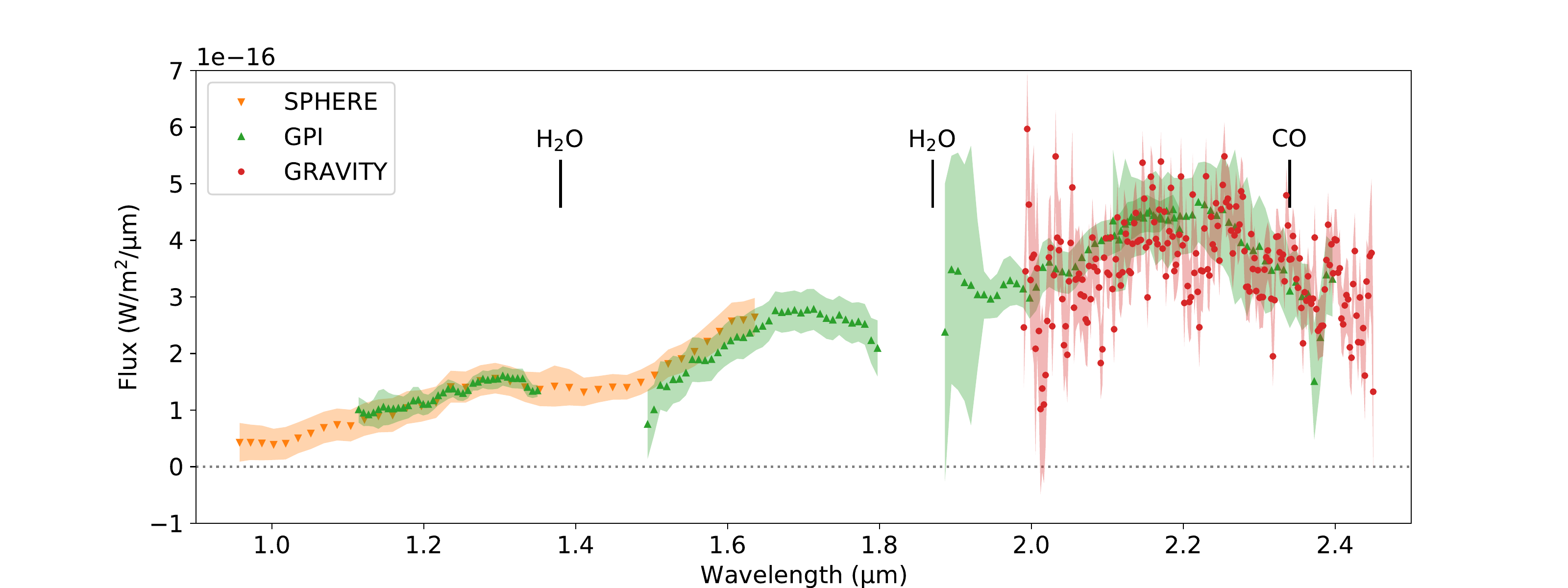}
\caption{Combined GRAVITY K-band spectrum of HD~206893~B together with the SPHERE Y--H-band spectrum from \citet{delorme2017} and the GPI J, H, K1, and K2-band spectra from \citet{ward-duong2020}. The shaded regions highlight the 1--$\sigma$ confidence intervals. For reference, absorption bands of water and carbon monoxide are indicated.}
\label{fig:spectra}
\end{figure*}

Apart from the two astrometric data points, we also obtain two K-band spectra at a resolution of $R \sim 500$ for HD~206893~B, one for each epoch of GRAVITY data. These two spectra are consistent with each other and we combine them into a single final spectrum, accounting for the covariances, shown in Figure~\ref{fig:spectra} together with the SPHERE spectrum from \citet{delorme2017} and the GPI spectra from \citet{ward-duong2020}.
In Sections~\ref{sec:atmospheric_model_fitting} and~\ref{sec:spectral_retrieval}, we use these spectra for atmospheric model fitting and spectral retrievals of HD~206893~B, respectively. To constrain the fits between $3.5$--$5~\text{\textmu m}$, we supplement the spectra with photometry of HD~206893~B from the literature (cf. Table~\ref{tab:hd206893b_photometry}).

\subsection{Atmospheric model fitting}
\label{sec:atmospheric_model_fitting}

\begin{table*}
\caption{Prior boundaries for the atmospheric model grids used in this work.}
\label{tab:prior_boundaries}
\centering
\begin{tabular}{c|cccccc}
\hline\hline
Model & $T_\text{eff}$ & $\log g$ & $[\text{Fe/H}]$ & C/O & $R$ \\
& (K) & & (dex) & & ($\text{R}_\text{Jup}$) \\
\hline
BT-Settl-CIFIST & 1000--2000 & 2.5--5.5 & -- & -- & 0.8--6.0 \\
DRIFT-PHOENIX & 1000--2000 & 3.0--5.5 & -0.6--0.3 & -- & 0.8--6.0 \\
Exo-REM & 1000--2000 & 3.5--4.5 & -0.5--0.5 & 0.3--0.75 & 0.8--6.0 \\
\hline
\multicolumn{6}{l}{\textbf{Notes.} \parbox[t]{10 cm}{The boundaries for the effective temperature and the radius were chosen based on previous works on HD~206893~B, while those for the other parameters exploit the maximum allowed range.}}
\end{tabular}
\end{table*}

\begin{table*}
\caption{Atmospheric parameters inferred for HD~206893~B using grid retrievals. The plain models are unmodified whereas the dusty models include additional extinction by high-altitude dust clouds made of enstatite grains. The values state the 68\% confidence intervals around the median.}
\label{tab:atmospheric_parameters}
\centering
\begin{tabular}{c | c c c c c c c c c | c}
\hline\hline
Model & $T_\text{eff}$ & $\log g$ & [Fe/H] & C/O & $R$ & $M$ & $a_\text{mean}$ & $\sigma_a$ & $A_V$ & $\chi_\text{red}^2$ \\
& (K) & & (dex) & & ($\text{R}_\text{Jup}$) & ($\text{M}_\text{Jup}$) & ($\text{\textmu m}$) & & (mag) & \\
\hline
Plain \\
\hline
BT & $1600^{+1}_{-1}$ & $3.50^{+0.00}_{-0.01}$ & -- & -- & $0.98^{+0.01}_{-0.01}$ & $1.17^{+0.02}_{-0.02}$ & -- & -- & -- & 0.957 \\
DP & $1431^{+9}_{-9}$ & $5.14^{+0.20}_{-0.13}$ & $0.27^{+0.02}_{-0.04}$ ($\uparrow$) & -- & $1.20^{+0.03}_{-0.03}$ & $77^{+45}_{-21}$ & -- & -- & -- & 0.841 \\
ER & $1049^{+2}_{-4}$ & $3.50^{+0.00}_{-0.00}$ ($\downarrow$) & $0.49^{+0.01}_{-0.01}$ ($\uparrow$) & $0.65^{+0.00}_{-0.00}$ & $2.32^{+0.02}_{-0.02}$ & $6.58^{+0.11}_{-0.10}$ & -- & -- & -- & 1.024 \\
\hline
Dusty \\
\hline
BT & $1589^{+13}_{-22}$ & $3.83^{+0.38}_{-0.14}$ & -- & -- & $1.17^{+0.07}_{-0.05}$ & $3.55^{+5.67}_{-1.08}$ & $0.33^{+0.05}_{-0.06}$ & $1.30^{+0.15}_{-0.12}$ & $1.99^{+0.44}_{-0.50}$ & 0.751 \\
DP & $1444^{+12}_{-11}$ & $5.02^{+0.14}_{-0.14}$ & $0.26^{+0.03}_{-0.05}$ ($\uparrow$) & -- & $1.75^{+0.25}_{-0.16}$ & $123^{+56}_{-35}$ & $2.29^{+0.11}_{-0.10}$ & $1.17^{+0.07}_{-0.05}$ & $0.82^{+0.23}_{-0.18}$ & 0.774 \\
ER & $1347^{+6}_{-7}$ & $3.55^{+0.06}_{-0.04}$ ($\downarrow$) & $0.06^{+0.09}_{-0.07}$ & $0.75^{+0.00}_{-0.01}$ ($\uparrow$) & $2.03^{+0.08}_{-0.08}$ & $5.71^{+1.01}_{-0.65}$ & $0.34^{+0.04}_{-0.04}$ & $1.34^{+0.11}_{-0.08}$ & $2.87^{+0.36}_{-0.30}$ & 0.757 \\
\hline
\multicolumn{11}{l}{\textbf{Notes.} \parbox[t]{17 cm}{Arrows ($\uparrow$ or $\downarrow$) indicate if a parameter converges toward the upper or lower grid boundary.}}
\end{tabular}
\end{table*}

By combining the GRAVITY spectrum with SPHERE and GPI spectra and photometry available in the literature, we reach a broad spectral coverage from $\sim 1$--$5~\text{\textmu m}$. This spectral region contains absorption bands of water, carbon-monoxide, and methane and is broad enough to estimate the effective temperature, the radius, and the surface gravity of an object. We estimate these parameters for HD~206893~B by fitting its spectra and photometry with atmospheric model grids using \texttt{species}\footnote{\url{https://github.com/tomasstolker/species}} \citep{stolker2020}. There is a variety of atmospheric model grids for giant planets and brown dwarfs, all of them being slightly different in terms of underlying physics and complexity. However, all of them assume radiative-convective equilibrium to calculate the temperature structure of the atmosphere self-consistently. Here, we use three different grids: the BT-Settl-CIFIST grid \citep{allard2012}, the DRIFT-PHOENIX grid \citep{helling2008a} (which includes metallicity as an additional free parameter), and the Exo-REM grid \citep{baudino2015,charnay2018} (which includes both metallicity and C/O ratio as additional free parameters). All three grids include photospheric absorption by dust clouds, but with different approaches to calculate the cloud densities, grain size distributions and compositions. We bin the grids to the spectral resolution of the respective instrument, and use the spectra and filter curves to calculate the synthetic photometry and filter-weighted average flux, respectively. For all grid parameters, we use uniform priors whose boundaries are presented in Table~\ref{tab:prior_boundaries}. Our atmospheric model fits account for the covariances in the GRAVITY spectrum according to Section 2 of \citet{greco2016}. While fits to the GRAVITY and the SPHERE spectra alone show good photometric agreement between the two, there seems to be a significant offset between the GPI H-band and the SPHERE spectrum, which may indicate a systematic error in the absolute flux calibration. Given that the SPHERE spectrum agrees well with the GPI J-band spectrum, we decided to fit a separate scaling parameter to each of the GPI spectra while keeping the GRAVITY and the SPHERE spectra fixed in order to preserve the extremely red color of HD~206893~B. Then, we infer the posterior distribution of the model parameters with nested sampling using \texttt{PyMultiNest}\footnote{\url{https://github.com/JohannesBuchner/PyMultiNest}} \citep{buchner2014,feroz2009,feroz2019}.

Several hypotheses have been put forward to explain the extreme redness of HD~206893~B, most notably extinction by local dust, either extra-photospheric or in the form of a circumplanetary disk by \citet{milli2017}, \citet{delorme2017}, and \citet{ward-duong2020}. Other possibilities like reddening by interstellar dust or extinction by the debris disk could be mostly ruled out. \citet{ward-duong2020} could not find any significant interstellar extinction toward the host star based on its photometry, and we can confirm this finding by visual inspection of stellar Ca-lines in high-resolution spectra of HD~206893~A (A.-M.~Lagrange, private communication). It seems unlikely that there is an interstellar dust cloud that is obscuring HD~206893~B but not its host star, which is separated by only $\sim 250~\text{mas}$. Moreover, the debris disk of the system would need to be unrealistically optically thick ($\tau \sim 1.7$) to explain the observed reddening, even if viewed edge-on \citep{ward-duong2020}. Therefore, extinction by local dust is the most plausible explanation for the extremely red color of HD~206893~B, and we include additional extinction by high-altitude dust clouds made of crystalline enstatite ($\text{MgSiO}_3$) grains in our atmospheric model fits. Since other dust species such as forsterite, corundum, and iron predict similar extinction curves for grain sizes between $0.1$--$1~\text{\textmu m}$ \citep[e.g.,][]{ward-duong2020}, we only consider enstatite grains for simplicity here. These grains are described by a log-normal size distribution
\begin{equation}
    \frac{dn}{da} = \frac{N}{a\sqrt{2\pi}\ln{\sigma_a}}\exp{\left(-\frac{\ln^2{(a/a_\text{mean})}}{2\ln^2{\sigma_a}}\right)},
\end{equation}
where $n$ is the number concentration of grains smaller than $a$, $N$ is the total number concentration of grains, $a$ is the grain size, $a_\text{mean}$ is the geometric mean grain size, and $\sigma_a$ is the grain size geometric standard deviation \citep[which is dimensionless,][]{ackerman2001}. A log-uniform prior between $0.1$--$10~\text{\textmu m}$ is used for $a_\text{mean}$ and a uniform prior between $1.1$--$10$ is used for $\sigma_a$. Smaller grains would grow by condensation within timescales of less than a second and are therefore not considered \citep{charnay2018}. Then, we compute the extinction cross-section $\sigma_{\text{ext},\lambda}(a_\text{mean},\sigma_a)$ using \texttt{PyMieScatt}\footnote{\url{https://github.com/bsumlin/PyMieScatt}} \citep{sumlin2018} and scale the flux of the default spectra $F_\text{default}$ to that of the reddened spectra
\begin{equation}
    F_\text{red} = F_\text{default} \cdot 10^{\left(-\frac{A_V}{2.5}\frac{\sigma_{\text{ext},\lambda}}{\sigma_{\text{ext},\text{V}}}\right)},
\end{equation}
where $A_V$ is the extinction in the Bessel V-band, another free parameter with a uniform prior between $0$--$5~\text{mag}$, and $\sigma_{\text{ext},\text{V}}$ is the extinction cross-section averaged over the Bessel V-band. In total, our enstatite dust model has three free parameters ($a_\text{mean}$, $\sigma_a$, and $A_V$), which are inferred along with the parameters of the atmospheric model grids.

\begin{figure*}
\centering
\includegraphics[width=\textwidth]{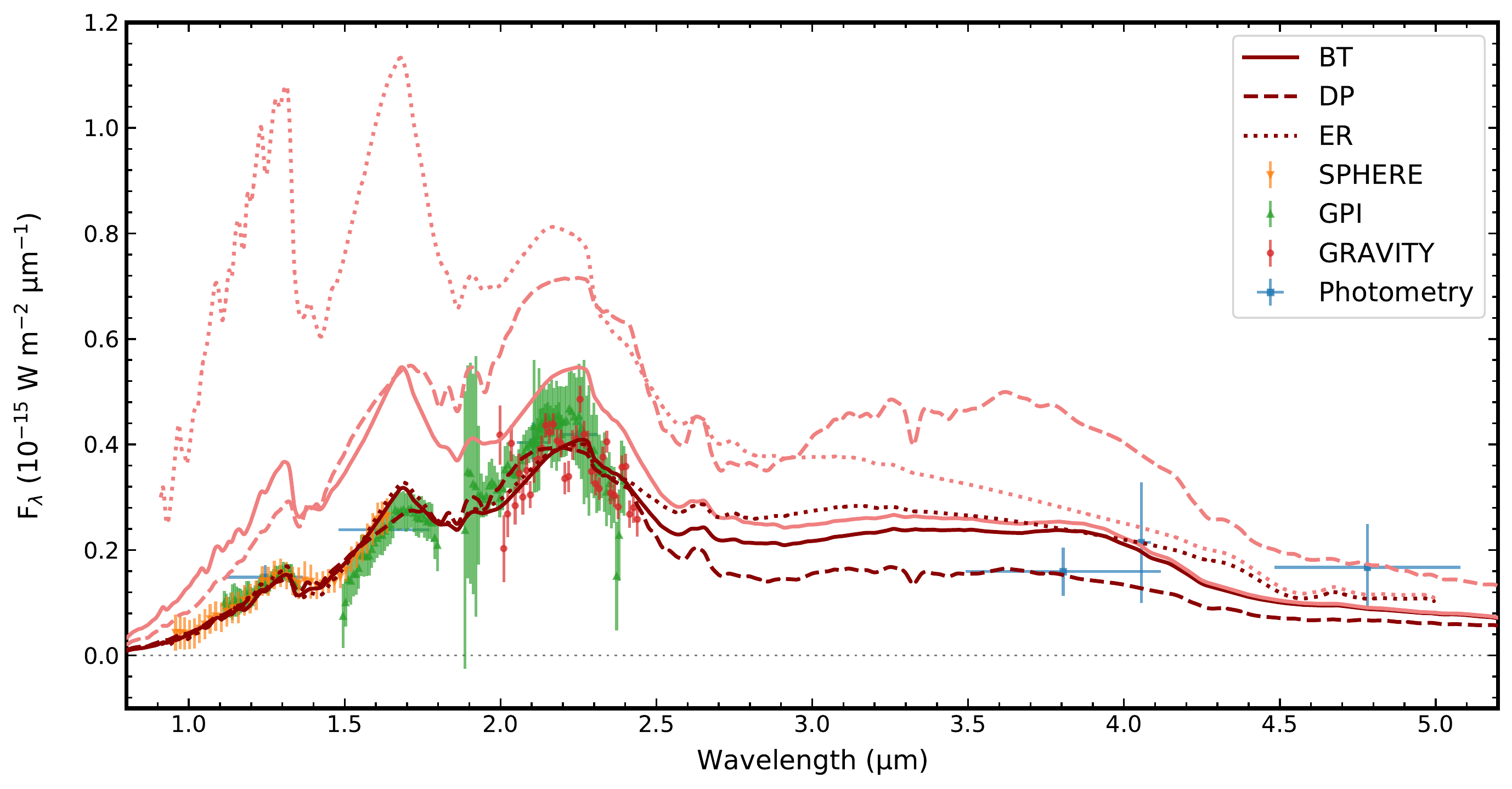}
\caption{Atmospheric models fitted to the observed spectra and photometry of HD~206893~B shown in the background. Dark red lines show the best fit dusty models that include additional extinction by high-altitude dust clouds made of enstatite grains and light red lines show the exact same models before applying the extinction (i.e., without dust). The GPI spectra are not to be taken to face value since they are rescaled during each of the fits. BT = BT-Settl-CIFIST, DP = DRIFT-PHOENIX, and ER = Exo-REM.}
\label{fig:spectrum}
\end{figure*}

\begin{figure*}
\centering
\includegraphics[width=\textwidth]{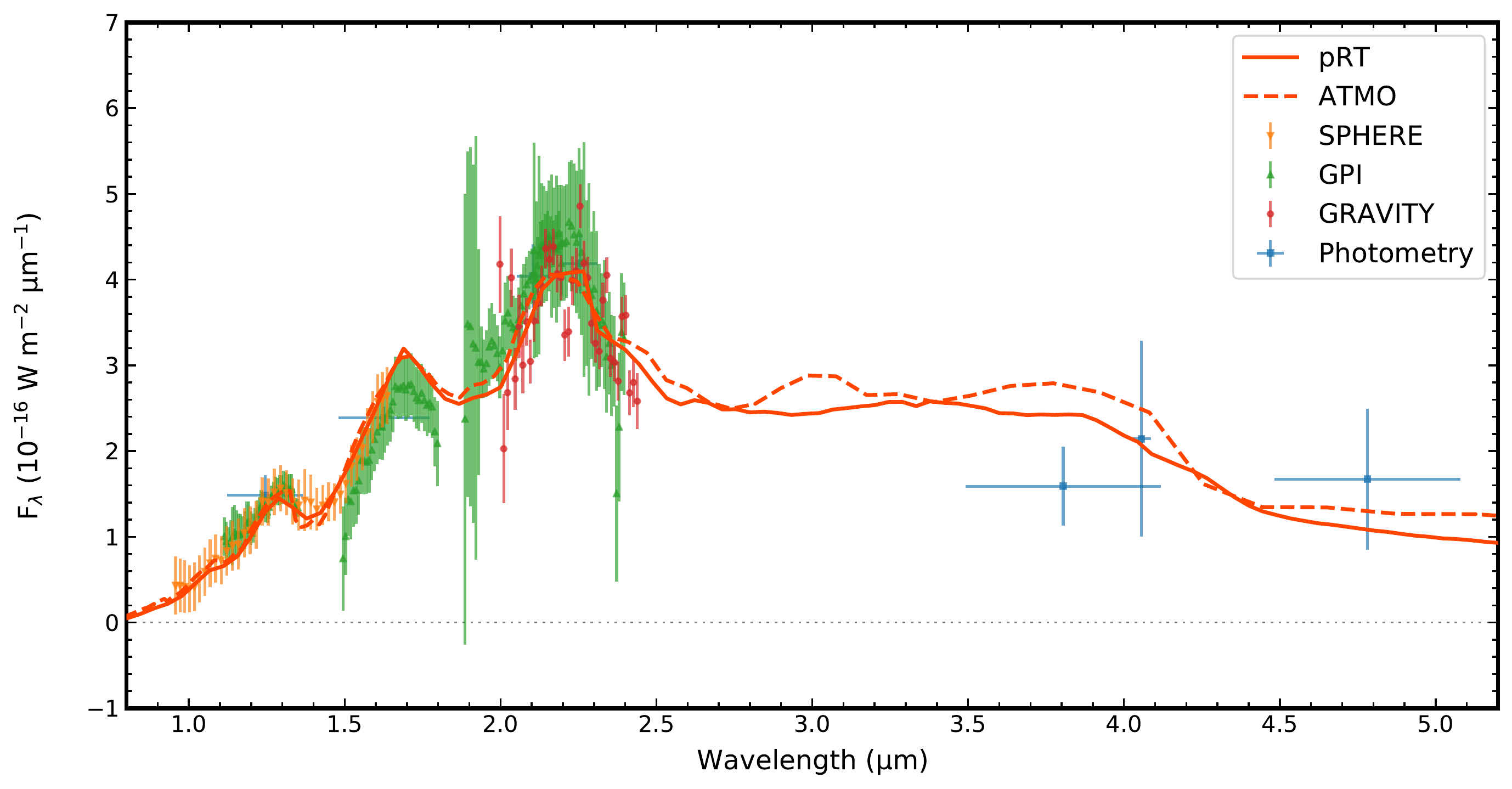}
\caption{Retrieved spectra of HD~206893~B with the observed spectra and photometry shown in the background. The GPI spectra are not to be taken to face value since they are rescaled during the retrieval with \texttt{petitRADTRANS} (\texttt{pRT}).}
\label{fig:spectrum_retr}
\end{figure*}

Table~\ref{tab:atmospheric_parameters} summarizes the atmospheric parameters obtained for HD~206893~B based on the three different atmospheric model grids without (``plain'') and with (``dusty'') additional extinction. The inferred effective temperatures are very similar to those obtained by \citet{delorme2017} and \citet{ward-duong2020}, but the surface gravities confirm the trend observed by \citet{ward-duong2020}, namely that the H and K-band spectra prefer lower surface gravities than those obtained by \citet{delorme2017} for the SPHERE spectrum at shorter wavelengths. Overall, the parameters inferred from the plain models are spread over a wider range of parameter space than those inferred from the dusty models. Moreover, all dusty models fit the observed data better than the plain models since they have smaller $\chi_\text{red}^2$. This is not completely surprising given that the dusty models have three more free parameters for describing the additional extinction than the plain models. Most noticeably, for both the plain and the dusty models the DRIFT-PHOENIX (DP) grid predicts a significantly higher surface gravity and mass for HD~206893~B than the BT-Settl-CIFIST (BT) and Exo-REM (ER) grids. However, while the DP grid yields the best fit (i.e., the smallest $\chi_\text{red}^2$) for the plain models, it yields the worst fit (i.e., the highest $\chi_\text{red}^2$) for the dusty models.

Another striking difference between the DP grid and the BT and ER grids are the extinction parameters that they predict for the dusty models. While the BT and ER grids consistently prefer small grains with a geometric mean size of $\sim 0.33$--$0.34~\text{\textmu m}$ and a geometric standard deviation of $\sim 1.30$--$1.34$, the DP grid prefers large grains with a geometric mean size of $\sim 2.29~\text{\textmu m}$ and a geometric standard deviation of $\sim 1.17$. This is a difference in geometric mean grain size of almost an order of magnitude. We note that the DP grid uses a different cloud model than the BT and ER grids. With DP, gas is mixed to high altitudes where dust then forms and grows as it rains down. With BT and ER, the cloud model from \citet{ackerman2001} is used which assumes that the cloud particles are mixed from the cloud base upward. These fundamentally different approaches could cause the observed difference in predicted dust grain size. \citet{helling2008b} have further found that the dust to gas ratio predicted by the DRIFT model is larger than the one predicted by the Settl model at small pressures, where the mean grain size is below $1~\text{\textmu m}$. However, with DP the difference in $\chi_\text{red}^2$ with and without extinction is small, that is, DP with large grains (dusty) does not fit the data much better than DP without large grains (plain).

Figure~\ref{fig:spectrum} shows the best fit model spectra for the dusty models in dark red together with the NACO, SPHERE, GPI, and GRAVITY spectra and photometry of HD~206893~B. It is noteworthy that there are significant differences between the BT and ER grids and the DP grid regarding the depth of the $1.4~\text{\textmu m}$ water absorption feature and the morphology of the H- and K-band peaks. The triangular shaped H-band peak observed by GPI and fit well by the BT and ER grids (but not the DP grids) is typical for a young low-gravity object \citep[e.g.,][]{kirkpatrick2012,allers2013}. Moreover, they deviate significantly at longer wavelengths ($> 2.5~\text{\textmu m}$). There, the available NACO photometry is not precise enough to set meaningful constraints on the model parameters.

The exact same three dusty models are shown in light red, but before applying the additional extinction. Here, the difference in predicted grain size between the DP grid and the BT and ER grids becomes very clear. While for the BT and ER grids, the difference between unextinct (light red) and extinct (dark red) model spectrum decreases with increasing wavelength and approaches zero over the L and M-band, the extinction reaches its maximum near the L-band (where the wavelength is on the order of the geometric mean grain size) for the DP grid. Finally, we note that the ER grid predicts a significantly higher V-band extinction than the BT grid, despite the very similar grain size parameters. This is the case because the ER grid converges toward a significantly larger radius if compared to the BT grid, resulting in a higher bolometric luminosity and therefore requiring higher extinction.

It is also noteworthy that in the absence of additional dust grains (i.e., the plain models), the metallicity is found to hit the upper bound of the DP and ER grids (i.e., significantly super-solar metallicity). It has been observed before that a high metallicity facilitates the formation of dust grains and might be an explanation for the unusually red L dwarf population \citep{looper2008,stephens2009,gizis2012,marocco2014}. A slight anticorrelation between metallicity and dust extinction ($A_V$) for the dusty ER grid (cf. Figure~\ref{fig:posterior_ER_dust}) supports the finding that higher metallicity causes a redder color. However, for the dusty DP grid, where the metallicity also hits the upper bound, we cannot identify such a correlation.

\subsection{Spectral retrieval}
\label{sec:spectral_retrieval}

The atmospheric model fits in Section~\ref{sec:atmospheric_model_fitting} have shown that HD~206893~B is difficult to explain with currently available grids and often drives parameters such as the surface gravity or the metallicity to the grid boundaries in order to mimic an extremely red color. Spectral retrievals are better suited for exploring a wide range of parameters and multi-species gas opacities \citep[e.g.,][]{ward-duong2020}. We thus perform spectral retrievals with \texttt{petitRADTRANS} \citep{molliere2019,molliere2020} and \texttt{ATMO} \citep{tremblin2015,tremblin2016} on the same spectra and photometry of HD~206893~B that we also used for the atmospheric model fits in Section~\ref{sec:atmospheric_model_fitting}.

\begin{table*}
\caption{Atmospheric parameters inferred for HD~206893~B using free retrievals.}
\label{tab:retrieval_parameters}
\centering
\begin{tabular}{c | c c c c c c c c}
\hline\hline
Model & $T_\text{eff}$ & $\log g$ & [Fe/H] & [C/C$_\odot$] & [O/O$_\odot$] & C/O & $R$ & $M$ \\
& (K) & & (dex) & (dex) & (dex) & & ($\text{R}_\text{Jup}$) & ($\text{M}_\text{Jup}$) \\
\hline
\texttt{petitRADTRANS} & $1216^{+13}_{-17}$ & $2.87^{+0.63}_{-0.47}$ & $1.62^{+0.35}_{-0.41}$ & -- & -- & $0.82^{+0.04}_{-0.19}$ & $1.70^{+0.05}_{-0.05}$ & $0.83^{+2.71}_{-0.55}$ \\
\texttt{ATMO} & $1113^{+51}_{-52}$ & $2.72^{+0.24}_{-0.10}$ & $0.74^{+0.16}_{-0.21}$ & $1.45^{+0.10}_{-0.14}$ & $1.24^{+0.09}_{-0.13}$ & $0.90^{+0.03}_{-0.04}$ & $2.23^{+0.09}_{-0.09}$ & $1.07^{+0.82}_{-0.24}$ \\
\hline
\end{tabular}
\end{table*}

\begin{figure}
\centering
\includegraphics[width=\columnwidth]{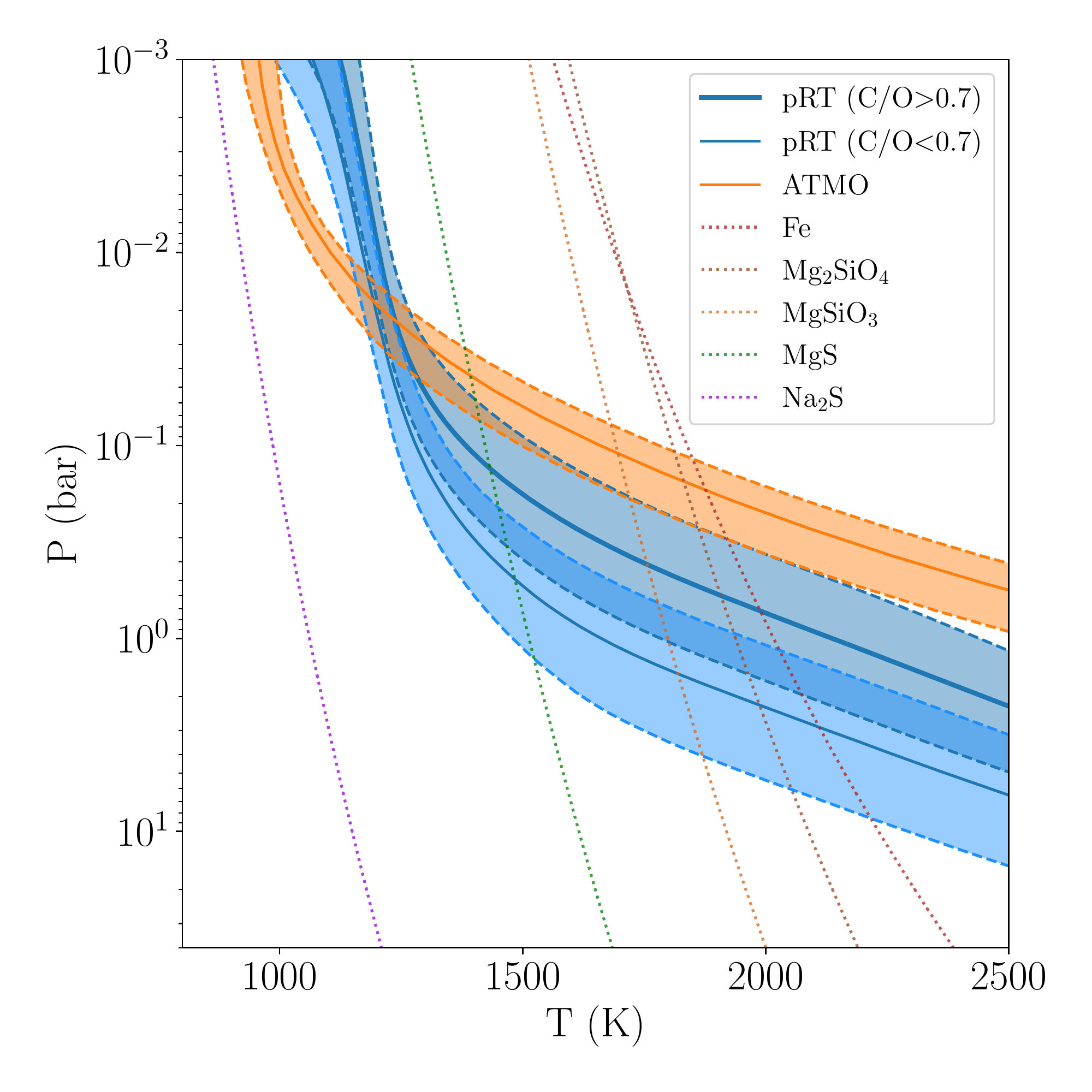}
\caption{Pressure-temperature profiles of the atmosphere of HD~206893~B retrieved with \texttt{petitRADTRANS} (\texttt{pRT}) and \texttt{ATMO}. The dashed lines show the 1--$\sigma$ confidence intervals. With \texttt{pRT} we obtain a bimodal posterior for the C/O ratio and the quenching pressure, and the $P$--$T$ profile for each of the two solutions is shown separately (thick blue line C/O > 0.7, thin blue line C/O < 0.7). All models probe approximately 0.01 to 0.1 bar pressure levels. Dotted lines indicate the condensation curves of various species calculated at 10$\times$ solar abundance \citep{visscher2010}.}
\label{fig:retrieval_pTs}
\end{figure}

\subsubsection{\texttt{petitRADTRANS}}

For the spectral retrieval with \texttt{petitRADTRANS} (\texttt{pRT}), we follow mostly the implementation including scattering clouds as described in \citet{molliere2020} for the case of HR~8799~e. We briefly summarize that the P-T structure is parameterized into three different regions. Specifically, we use free temperature nodes at high altitudes, the Eddington approximation for the photospheric region, and a moist adiabat for the deep, radiative-convective part of the atmosphere. Gas abundances are assumed to be in chemical equilibrium, but with an additional parameter for a quenching pressure above which the CO, CH$_4$, and H$_2$O abundances are fixed. We include CO, H$_2$O, CH$_4$, NH$_3$, CO$_2$, H$_2$S, Na, K, PH$_3$, VO, TiO, FeH as molecular and atomic line species plus Rayleigh scattering and collision induced absorption (CIA) of H$_2$ and He. Given the similar temperature to HR~8799~e, we only consider cloud condensates composed of MgSiO$_3$ and Fe.

Since HD~206893~B is thought to have an unusually cloudy atmosphere, we use the cloud optical depth in the photospheric region as free parameter instead of the mass fractions themselves. The cloud mass fraction above the cloud base, $X_\mathrm{cloud}$, is parameterized as \citep{molliere2020}
\begin{equation}
    X_\mathrm{cloud}(P) = X_{\mathrm{cloud},0}\left(\frac{P}{P_\mathrm{base}}\right)^{f_\mathrm{sed}},
\end{equation}
where $X_{\mathrm{cloud},0}$ is the cloud mass fraction at the cloud base, $P_\mathrm{base}$ is the pressure at the cloud base, and $f_\mathrm{sed}$ is the settling parameter (assumed to be the same for MgSiO$_3$ and Fe). The cloud parameter, $\tau_\mathrm{cloud}$, is then defined as the optical depth at the $\tau = 1$ pressure of the gas-only atmosphere (i.e., neglecting the clouds). To ensure a quasi-physical solution for the cloud properties, we reject samples for which
\begin{equation}
    X_{\mathrm{cloud},0} > 2X_\mathrm{eq}(1+f_\mathrm{sed}),
\end{equation}
where $X_\mathrm{eq}$ is the equilibrium mass fraction at the cloud base as calculated from the elemental abundances and [Fe/H]. This expression stems from requiring that the surface density of the cloud does not exceed the surface density of condensing species in the atmospheric column above the cloud base, where we allowed for an additional factor of two to slightly relax this condition.

Similar to the atmospheric model fits in Section~\ref{sec:atmospheric_model_fitting}, we use a nested sampling algorithm \citep{feroz2009,buchner2014} to sample the posterior distributions of the 20 free parameters with 4000 live points. The model spectra are smoothed to the respective instrument resolution before they are resampled to the data wavelengths, and a separate scaling factor is fitted to each of the GPI spectra to account for systematic calibration offsets. The results of the retrieval are presented in Table~\ref{tab:retrieval_parameters} and the retrieved spectrum and P-T profile are shown in Figures~\ref{fig:spectrum_retr} and~\ref{fig:retrieval_pTs}, respectively. We note that, for clarity, the scaling factors have not been applied when plotting the spectra, to be able to show the best fit models of \texttt{pRT} and \texttt{ATMO} (described below), the latter of which did not retrieve scaling factors.

\subsubsection{\texttt{ATMO}}

Another spectral retrieval was also performed with \texttt{ATMO}, a 1D-2D radiative transfer model for planetary atmospheres \citep{tremblin2015, tremblin2016}. More comprehensive descriptions of the model can be found in \citet{amundsen2014}, \citet{drummond2016}, \citet{goyal2018}, and \citet{phillips2020}. The retrieval aspects of \texttt{ATMO} have been used to fit transit and secondary eclipse data before \citep[e.g.,][]{evans2017,wakeford2017}, and are applied here on the HD~206893~B data.

We fit the data assuming chemical equilibrium, and include rainout \citep{goyal2020} to account for the depletion of gas phase species due to condensation. The total opacity of the gas mixture is computed using the correlated-$k$ approximation using the random overlap method with resorting and rebinning \citep{amundsen2014}. The $k$-coefficients and chemical equilibrium are calculated ``on the fly'' for each atmospheric layer, spectral band, and iteration such that the derived opacities are physically self-consistent with the P-T profile and chemical composition for each model evaluation in the retrieval. We include spectrally active species H$_2$-H$_2$ and H$_2$-He CIA opacities, as well as H$_2$O, CO$_2$, CO, CH$_4$, NH$_3$, K, TiO, VO, FeH, CrH, HCN, C$_2$H$_2$, H$_2$S, and H- (see \citealt{goyal2018} for further details). We fit for the elemental abundances of C and O separately, with the rest of the elements described by a single metallicity parameter, $\rm [Fe/H]$. 

Scattering and absorption effects from clouds on the spectrum were parameterized as follows. We include scattering from small particles using an enhanced Rayleigh-like scattering \citep{lecavelier2008} opacity, parameterized as
\begin{equation}
    \sigma(\lambda)_{\mathrm{haze}} = \delta_{\mathrm{haze}}\sigma_{\mathrm{0}}(\lambda/\lambda_0)^{\alpha_{\mathrm{haze}}},
\end{equation}
where $\sigma(\lambda)$ is the total scattering cross-section of the material, $\delta_{\mathrm{haze}}$ is an empirical enhancement factor, $\sigma_{\mathrm{0}}$ is the scattering cross section of molecular hydrogen at $0.35~\text{\textmu m}$, and $\alpha_{\mathrm{haze}}$ is a factor determining the wavelength dependence. Condensate cloud absorption is fit separately, and is assumed to have a gray wavelength dependence calculated as 
\begin{equation}
    \kappa(\lambda)_{\mathrm{cloud}} = \delta_{\mathrm{cloud}}\kappa_{\mathrm{H2}}, 
\end{equation}
where $\kappa(\lambda)_{\mathrm{cloud}}$ is the ``cloud'' absorption opacity, $\delta_{\mathrm{cloud}}$ is an empirical factor governing the strength of the gray scattering, and $\kappa_{\mathrm{H2}}$ is the scattering opacity due to H$_{2}$ at $0.35~\text{\textmu m}$. $\sigma(\lambda)_{\mathrm{haze}}$ and $\kappa(\lambda)_{\mathrm{cloud}}$ are added to the total gaseous scattering and absorption, respectively, with a further parameter specifying the pressure level at the top of the gray cloud. To parameterize the P-T profile, we use an analytic radiative equilibrium model by \cite{guillot2010}. Again, we use nested sampling to sample the posterior distribution \citep{feroz2009}, fitting for a total of twelve free parameters. A flux scaling for calibration offsets was not applied, however. The results of the retrieval are presented in Table~\ref{tab:retrieval_parameters} and the retrieved spectrum and P-T profile are shown in Figures~\ref{fig:spectrum_retr} and~\ref{fig:retrieval_pTs}, respectively.

\subsection{Evolutionary tracks}
\label{sec:evolutionary_tracks}

\begin{figure*}
\centering
\includegraphics[width=0.49\textwidth]{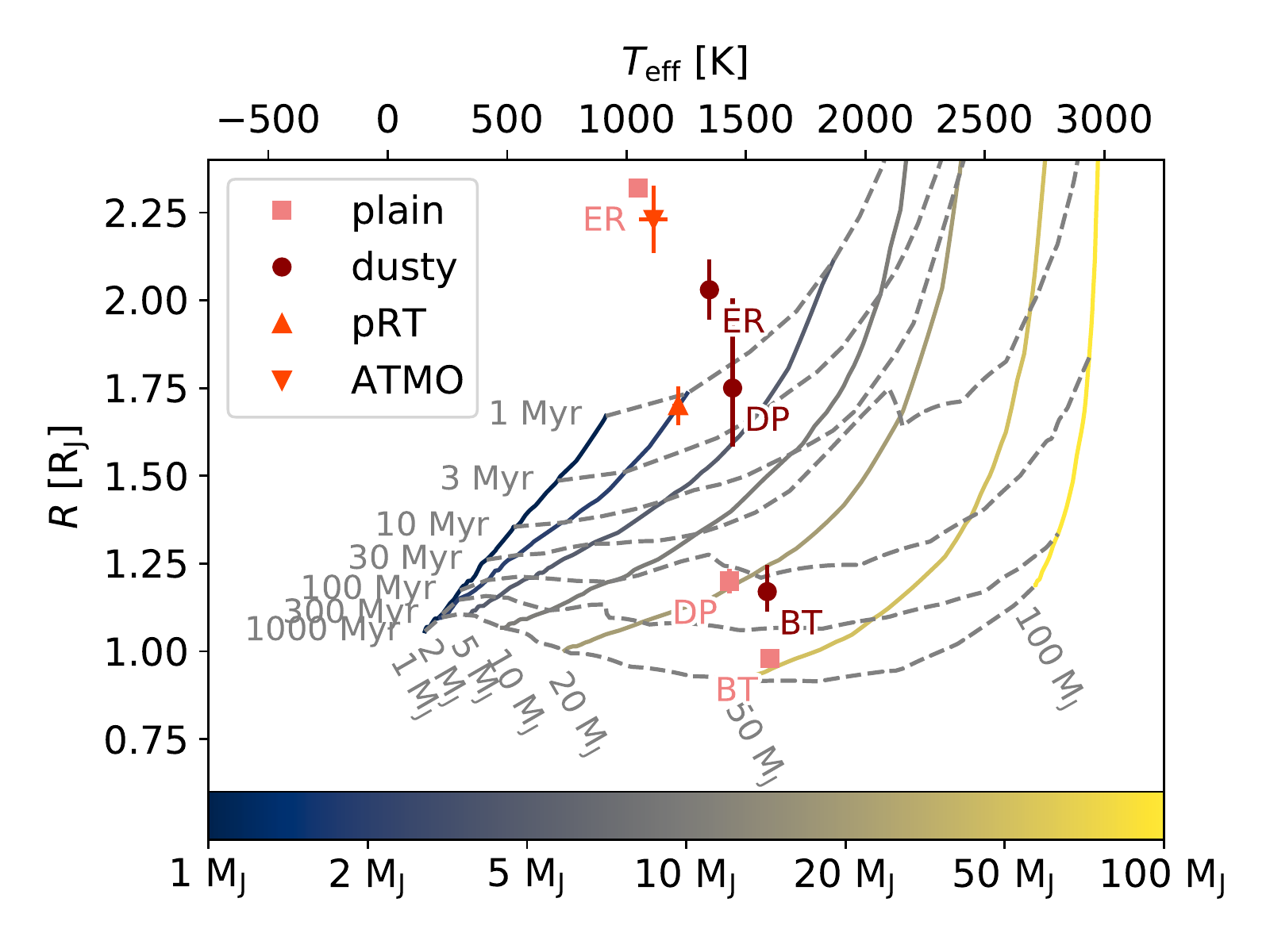}
\includegraphics[width=0.49\textwidth]{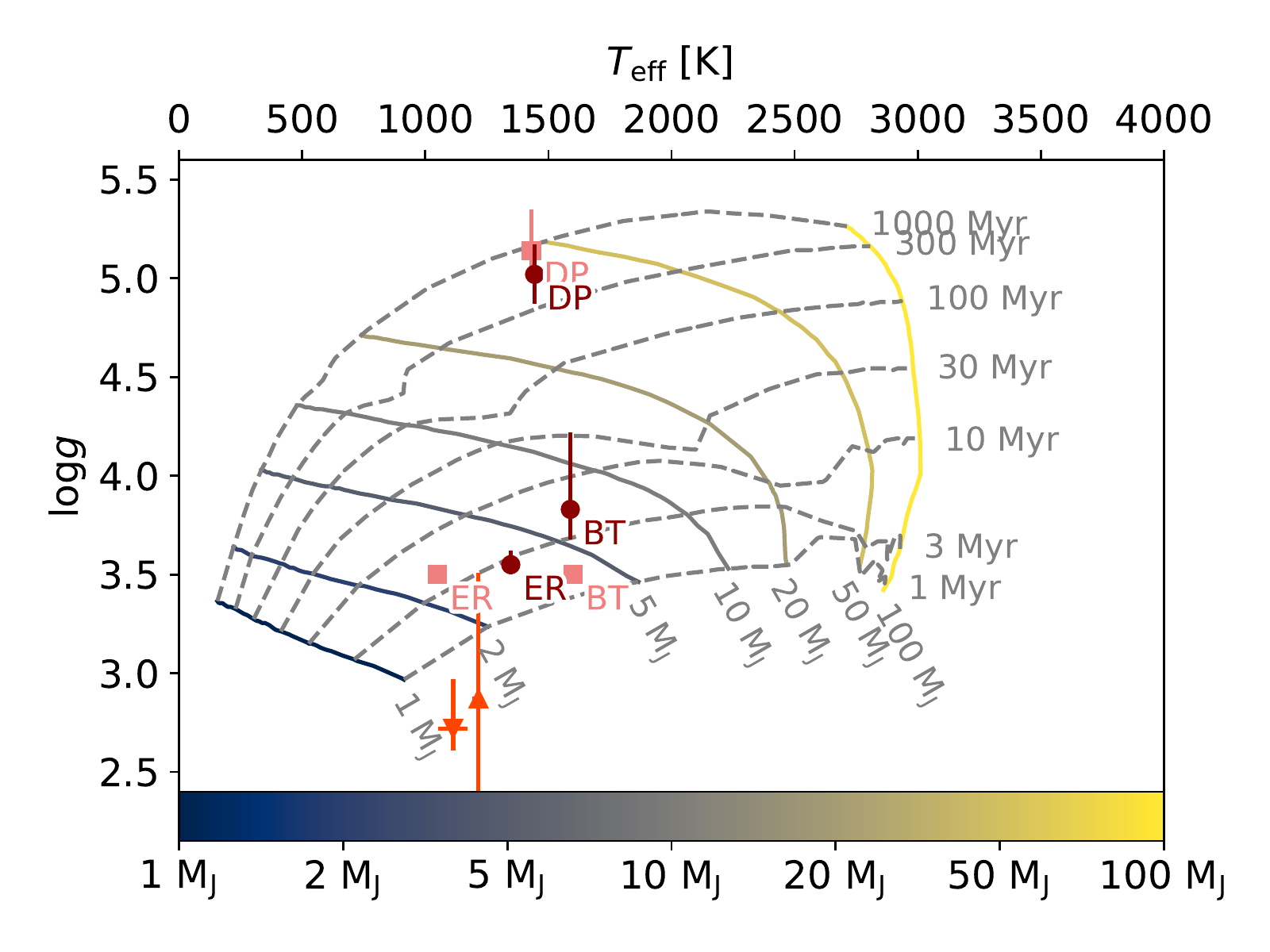}
\caption{Parameters inferred for HD~206893~B from atmospheric model fits and spectral retrievals compared to AMES-Cond evolutionary tracks. Light red points show the best fit parameters for the plain models and dark red points show the best fit parameters for the dusty models including additional extinction by high-altitude dust clouds made of enstatite grains. The evolutionary tracks are shown for objects with exactly those masses printed below the colorbar. Curves of constant age are shown in dashed gray. BT = BT-Settl-CIFIST, DP = DRIFT-PHOENIX, ER = Exo-REM, and \texttt{pRT} = \texttt{petitRADTRANS}.}
\label{fig:evotrack}
\end{figure*}

Similar to \citet{delorme2017}, we compare our best fit atmospheric parameters to evolutionary tracks for giant planets and brown dwarfs to ensure that they correspond to physically plausible substellar companions. Therefore, we use the AMES-Cond evolutionary tracks \citep{baraffe2003}. Figure~\ref{fig:evotrack} shows these tracks for objects of different masses from $1$--$100~\text{M}_\text{Jup}$ in the radius versus effective temperature and surface gravity versus effective temperature planes. The best fit atmospheric parameters are overplotted in light red (plain models), dark red (dusty models), and orange red (free retrievals).

The most significant outliers are the plain ER grid and the \texttt{ATMO} retrieval, which predict unexpectedly large radii of $2.32^{+0.02}_{-0.02}~\text{R}_\text{Jup}$ and $2.23^{+0.09}_{-0.09}~\text{R}_\text{Jup}$ for a relatively cool ($1049^{+2}_{-4}~\text{K}$ and $1113^{+51}_{-52}~\text{K}$) and low-mass ($6.58^{+0.11}_{-0.10}~\text{M}_\text{Jup}$ and $1.07^{+0.82}_{-0.24}~\text{M}_\text{Jup}$) object, respectively. With the additional extinction, the radius of the dusty ER grid decreases and the effective temperature increases, leading to an object that is roughly consistent between the atmospheric models and the evolutionary tracks with an extremely young ($< 10~\text{Myr}$) planetary-mass ($< 5~\text{M}_\text{Jup}$) companion, given the uncertainties on the surface gravity and the radius from the atmospheric model fits. In the same parameter range, the \texttt{pRT} retrieval can be found, but with a slightly smaller radius and an even lower surface gravity. We note, however, that the dusty ER grid converges toward the lower boundary of the surface gravity ($\sim 3.5$). Other inconsistent parameters are predicted by both the plain BT and dusty DP grids. While the low surface gravity predicted by the plain BT grid is consistent with an extremely young object ($< 3~\text{Myr}$), its predicted small radius is consistent with a rather old object ($> 300~\text{Myr}$). The same is observed the other way around for the dusty DP grid. Here, the predicted high surface gravity is consistent with a rather old object and the predicted large radius is consistent with an extremely young object. However, the dusty BT and plain DP grids predict objects that are roughly consistent between the atmospheric parameters and the evolutionary tracks. The dusty BT grid suggests a moderately young ($\sim 3$--$300~\text{Myr}$) object somewhere between $\sim 5$--$30~\text{M}_\text{Jup}$. The plain DP grid suggests a rather old ($\sim 100$--$1000~\text{Myr}$) object somewhere between $\sim 15$--$75~\text{M}_\text{Jup}$. We note that such objects are also in agreement with the age and mass predicted for HD~206893~B by \citet{delorme2017}.

Overall, we find that the BT and ER grids require additional extinction in order to predict physically plausible objects. This is different for the DP grid, which becomes unphysical when additional extinction is included. This could also be related to the much larger grain sizes predicted by the DP grid if compared to the BT and ER grids, which are not expected for high-altitude dust clouds \citep{hiranaka2016}. Moreover, we find that the three different atmospheric model grids predict objects that populate a wide range of parameter space in age and mass. The (dusty) ER grid predicts an extremely young ($< 10~\text{Myr}$) object of $< 5~\text{M}_\text{Jup}$, the (dusty) BT grid predicts a moderately young ($\sim 3$--$300~\text{Myr}$) object of $\sim 5$--$30~\text{M}_\text{Jup}$, and the (plain) DP grid predicts a rather old ($100$--$1000~\text{Myr}$) object of $\sim 15$--$75~\text{M}_\text{Jup}$.

\subsection{Color-magnitude diagram}
\label{sec:color-magnitude_diagram}

\begin{figure*}
\centering
\includegraphics[width=\textwidth]{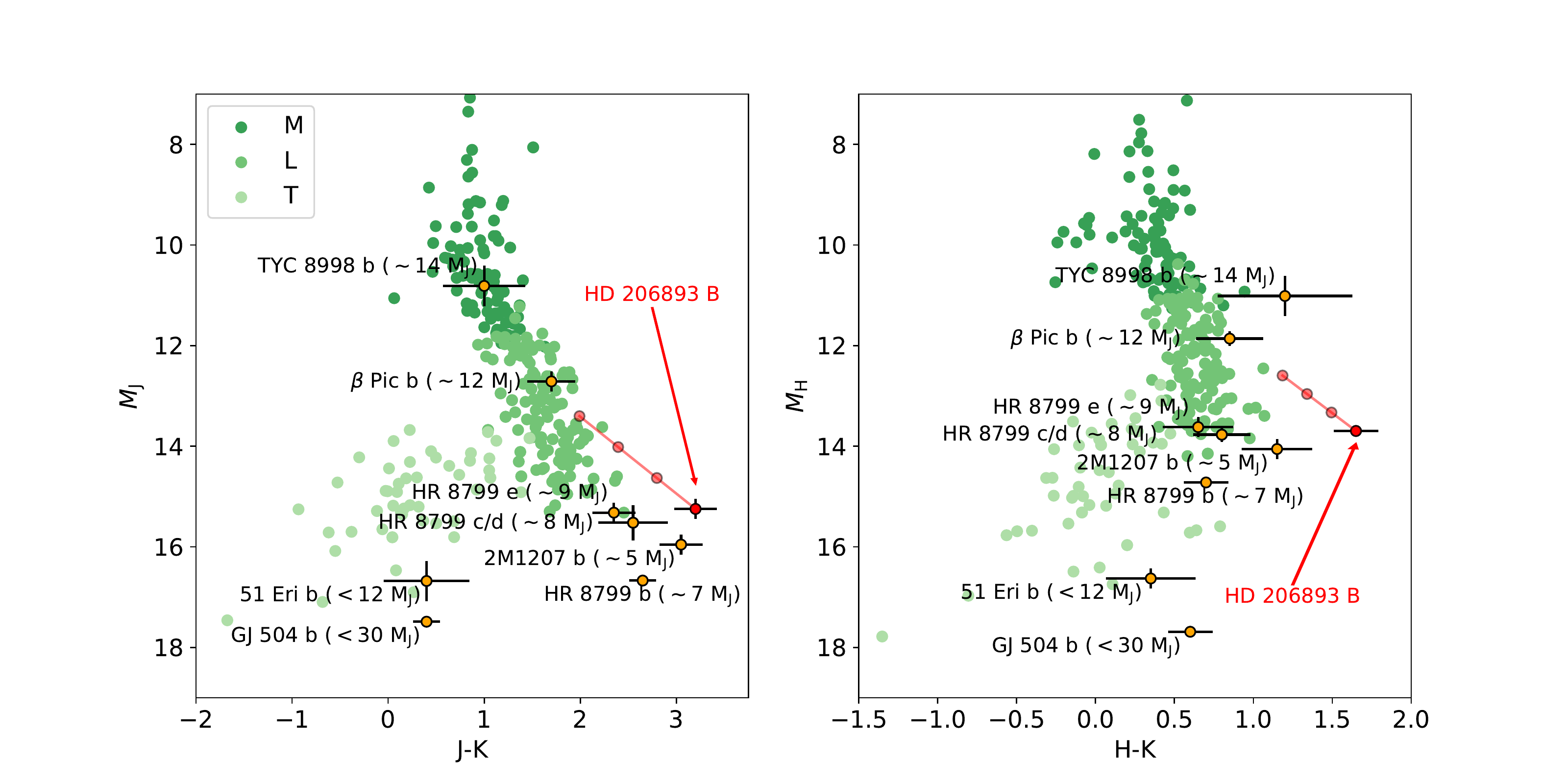}
\caption{J-K (left) and H-K (right) color-magnitude diagrams showing HD~206893~B in red together with the reddening vector of our best fit extra-photospheric enstatite dust model for different V-band extinctions in light red. Other known planetary-mass objects are shown in orange and M, L, and T-dwarfs from the SpeX Prism Spectral Libraries are shown in green. For the dust model, we assume $a_\text{mean} = 0.33~\text{\textmu m}$, $\sigma_a = 1.30$, and $A_V = 0$--$3$. The red points are in steps of 1~mag.}
\label{fig:colormag}
\end{figure*}

In a color-magnitude diagram, it can easily be seen that HD~206893~B is the reddest known substellar object \citep{milli2017,delorme2017,ward-duong2020}. Figure~\ref{fig:colormag} shows J-K and H-K color-magnitude diagrams of HD~206893~B and other known planetary-mass companions. For reference, M, L, and T-dwarfs from the SpeX Prism Spectral Libraries\footnote{\url{http://pono.ucsd.edu/~adam/browndwarfs/spexprism/library.html}} are shown in the background. The apparent magnitudes of the planetary-mass companions were taken from \citealt{delorme2017} (HD~206893~B), \citealt{currie2013} ($\beta$~Pic~b), \citealt{skemer2016} (GJ~504~b), \citealt{rajan2017} (51~Eri~b), \citealt{zurlo2016} (HR~8799~b, c, d, and e), \citealt{patience2012} (2M1207~b), and \citealt{bohn2020} (TYC~8998~b). Apparent magnitudes were converted to absolute magnitudes using distances (i.e., parallaxes) from SIMBAD\footnote{\url{http://simbad.u-strasbg.fr/simbad/}}.

The light red lines show the reddening vectors of our best fit enstatite dust model for different V-band extinctions. Here, we assume $a_\text{mean} = 0.33~\text{\textmu m}$ and $\sigma_a = 1.30$, consistent with the best fit dusty BT and ER models. We did not plot the reddening vector for the best fit dusty DP model because it corresponds to a physically implausible object (cf. Section~\ref{sec:evolutionary_tracks}). The shown reddening vectors extend from $A_V = 0$--$3$. For the best fit dusty BT and ER models, the predicted V-band extinction of $A_V \sim 2.0$ and $A_V \sim 2.9$, respectively, brings HD~206893~B back to the red end of the substellar main sequence, close to where the planetary-mass object $\beta$~Pic~b is located. This implies that HD~206893~B could indeed be a very dusty companion around a young moving group member, such as $\beta$~Pic~b, and marks another similarity between the HD~206893 and the $\beta$~Pic system besides the very similar system architecture. Finally, compared to the interstellar reddening law applied by \citet{ward-duong2020}, our enstatite dust model predicts a similar reddening slope in the $M_H$ versus H-K color-magnitude diagram while requiring smaller V-band extinction values of $\sim 2$--$3$ instead of $\sim 10$ in order to bring HD~206893~B back to the red end of the substellar main sequence (cf. their Figure~5). The existence of the CT~Cha companion \citep{schmidt2008} mentioned by \citet{ward-duong2020} and suffering from an extreme V-band extinction of $\sim 5$ magnitudes therefore puts the values of $A_V \sim 2$--$3$ obtained for HD~206893~B in a realistic regime.


\section{Discussion}
\label{sec:discussion}

\subsection{Astrometry}
\label{sec:astrometry}

In Section~\ref{sec:astrometric_analysis}, we infer the orbital parameters of HD~206893~B for two different scenarios. Scenario 1 is constrained by the data only and scenario 2 has an additional constraint on coplanarity with the debris disk of the system observed by \citet{milli2017} and \citet{marino2020}. By comparison with earlier works from \citet{grandjean2019} and \citet{ward-duong2020}, we find that the GRAVITY data resolves the degeneracy between a lower eccentricity, larger semimajor axis and a higher eccentricity, smaller semimajor axis orbit by preferring the latter of these in scenario 1. This is interesting because this orbital solution for HD~206893~B is only roughly aligned with respect to the debris disk of the system ($i_\text{m} < 34.4~\text{deg}$). \citet{marino2020} mention that a misalignment between the orbit of HD~206893~B and the debris disk of the system should be unlikely given its age of at least $50~\text{Myr}$ \citep{delorme2017}. They argue that HD~206893~B should align with the debris disk of the system due to secular interactions on timescales of only $\sim 10~\text{Myr}$. In scenario 2, where we enforce alignment between the orbit of HD~206893~B and the debris disk of the system, we clearly find that the lower eccentricity, larger semimajor axis orbit is preferred. This is in agreement with \citet{marino2020}, who obtained the same result without the additional GRAVITY data, and suggests that the GRAVITY data alone prefers a slight misalignment between the orbit of HD~206893~B and the debris disk of the system.

If this misalignment is confirmed by future GRAVITY observations, an explanation for it needs to be found. One possibility would be a significantly younger age ($< 50~\text{Myr}$) for the system. We note that such a young age would be in agreement with the age constraint set by comparing our best fit dusty BT and ER models and the spectral retrievals with evolutionary tracks (cf. Section~\ref{sec:evolutionary_tracks}). In addition, an age of $\sim 40~\text{Myr}$ would be expected according to \citet{torres2008} if the system was part of the Argus moving group, for which the probability is $\sim 61\%$ \citep{ward-duong2020}. We note, however, that the analysis of lithium and barium abundances points to an older age for the host star \citep{delorme2017}. The discrepancy between the predicted Argus moving group membership from Banyan Sigma \citep[$\sim 61\%$,][]{ward-duong2020} and Banyan II \citep[$\sim 13.5\%$,][]{delorme2017} can be attributed to the revised kinematics of the Argus association put forth by \citet{zuckerman2019}, though, and another recent and distinct Bayesian methodology by \citet{lee2019} also provides a membership probability of $\sim 63\%$. Another possibility would be dynamical interactions between HD~206893~B and the other putative companions predicted to exist in this system \citep[e.g.,][]{wu2011}. There is evidence for a second, massive ($\sim 15~\text{M}_\text{Jup}$) and close ($1.4$--$2.6~\text{au}$) companion interior to the orbit of HD~206893~B from radial velocity and \emph{Gaia} data \citep{grandjean2019} and a third $0.4$--$1.7~\text{M}_\text{Jup}$ companion further out responsible for carving the gap at $\sim 74~\text{au}$ in the debris disk observed with the Atacama Large Millimeter/Submillimeter Array \citep{marino2020}. However, a profound analysis of potential planet-planet interactions is beyond the scope of this work. Finally, we note that an eccentricity of $\sim 0.3$ for HD~206893~B preferred by the GRAVITY data is in better agreement with the eccentricity distribution of the brown dwarf population than that of wide-separation ($5$--$100~\text{au}$) giant planets, the latter of which show a preference for low eccentricities \citep[$\approx 0.05$--$0.25$,][]{bowler2020}. Still, sorting an individual companion into one or the other class of objects based on its eccentricity remains highly speculative, and as highlighted before, there might be other mechanisms responsible for an excitement of HD~206893~B's eccentricity, such as dynamical interactions with other companions in the system.

\subsection{Grid retrievals}
\label{sec:grid_retrievals}

In Section~\ref{sec:spectral_analysis}, we analyze the near-infrared spectrum of HD~206893~B. We obtain a dense spectral coverage between $1$--$2.5~\text{\textmu m}$ by combining data from SPHERE, GPI, and GRAVITY that we extend up to $\sim 5~\text{\textmu m}$ with VLT/NACO photometry from \citet{stolker2020}. While \citet{delorme2017} and \citet{ward-duong2020} found that currently available atmospheric models for giant planets and brown dwarfs fail to predict the extremely red color and the very shallow $1.4~\text{\textmu m}$ water absorption feature of HD~206893~B at the same time, they also compared the spectrum of HD~206893~B to those of other dusty, low-gravity or young M and L-dwarfs and found that none of these objects could reproduce all spectral features of HD~206893~B. Hence, both of these authors tried to reconcile the spectrum of HD~206893~B with those of other dusty, low-gravity or young M and L-dwarfs by including additional extinction by high-altitude dust clouds. Therefore, they explored a variety of dust species (forsterite, enstatite, corundum, and iron), grain sizes ($0.05$--$1~\text{\textmu m}$), and extinction values. They found that reddening by forsterite or enstatite grains with sizes between $0.27$--$0.50~\text{\textmu m}$ yields the best match to a very low-gravity L3-dwarf. Such a reddening by small ($< 1~\text{\textmu m}$) dust grains in the cool upper atmosphere has been suggested before by \citet{marocco2014} and \citet{hiranaka2016} to match the spectra of unusually red L-dwarfs with those of spectroscopic standards. They concluded that scattering by dust clouds with grain sizes between $1$--$100~\text{\textmu m}$ included in current atmospheric models is not sufficient to describe the peculiarly red L-dwarfs, which seem to require additional scattering by smaller grains in the cool upper atmosphere.

Here, we focused on a slightly different approach by adding extinction by high-altitude dust clouds directly to the atmospheric model grids. Since \citet{ward-duong2020} mention that all considered dust species predict similar extinction curves for grain sizes between $0.1$--$1~\text{\textmu m}$, we decided to only consider enstatite grains for simplicity here. Our dusty BT, DP, and ER grids fit the extremely red color as well as the very shallow $1.4~\text{\textmu m}$ water absorption feature of HD~206893~B well (cf. Figure~\ref{fig:spectrum}). However, both the plain and dusty DP grids fail to predict the pointy H-band peak observed with GPI\footnote{We note that since we allow for different scaling parameters for each of the GPI spectra during each of the fits, we cannot use them to obtain information about the absolute flux, but they are still useful to compare the spectral morphology between data and model.}. Such a triangular H-band peak is indicative of a low-gravity object for spectral types between M6--L7, although an older and dusty field brown dwarf with a similar spectral characteristic has been observed, too \citep{allers2013}. The H-band morphology of HD~206893~B is therefore at least suggestive of a dusty atmosphere, but likely also a low surface gravity as predicted by the BT and ER grids. Therefore, we obtain the best fits with our dusty BT and ER grids with $\chi^2 = 0.751$ and $0.757$, respectively. In agreement with the comparison between HD~206893~B and other dusty, low-gravity or young M and L-dwarfs by \citet{delorme2017} and \citet{ward-duong2020}, we find a grain size distribution with a geometric mean size of $\sim 0.3~\text{\textmu m}$ for these grids. Again, we note that the best fit dusty DP model deviates significantly from the best fit dusty BT and ER models at longer wavelengths ($> 2.5~\text{\textmu m}$) and better data, for example from the \emph{James Webb Space Telescope}, is required to obtain tighter constraints.

\subsection{Free retrievals}
\label{sec:free_retrievals}

The spectral retrievals, which explore a larger range of atmospheric parameters, yield spectra that are largely consistent with the best fit dusty BT and ER models regarding the spectral morphology between $1$--$2.5~\text{\textmu m}$ and the predicted flux at longer wavelengths (cf. Figure~\ref{fig:spectrum_retr}). The retrieved $T_\mathrm{eff}$ from the free retrievals (cf. Table~\ref{tab:retrieval_parameters}) is much below the values from the grid retrievals (cf. Table~\ref{tab:atmospheric_parameters}). This is the case because for the grid retrievals $T_\mathrm{eff}$ is computed for the interpolated spectrum without the additional extinction applied while for the free retrievals $T_\mathrm{eff}$ is obtained by integrating over the final spectrum. Both the free retrievals and the grid retrievals with BT and ER point to a low-gravity atmosphere, which, combined with the retrieved radii, interestingly points to a planetary-mass object. The metallicity and C/O ratio are super-solar in all cases and they are particularly enhanced according to the free retrievals. The high metallicity could also be suggestive of a low-mass object since an anticorrelation between mass and heavy element enrichment has been inferred in the atmospheres of the solar system ice and gas giants and exoplanets \citep{thorngren2016}, with planet formation models predicting [Fe/H] > 1 only for planets less massive than $\sim 1~\text{M}_\text{Jup}$ \citep[see Figure~3 of][]{mordasini2016}.

The retrieved pressure-temperature profiles from \texttt{pRT} and \texttt{ATMO} (cf. Figure~\ref{fig:retrieval_pTs}) show different gradients but their photospheres are located at similar pressures ($\sim 10$--$100~\text{mbar}$). The comparison with the condensation profiles in Figure~\ref{fig:retrieval_pTs} shows that MgS, MgSiO$_3$, Mg$_2$SiO$_4$, and Fe clouds are expected to be present in the atmosphere of HD~206893~B. The three latter species would condense out around 0.1~bar when considering the P-T profiles from \texttt{ATMO} although cloud opacities were parametrized during the retrieval without making an assumption about the cloud composition and only limited in their extent by the cloud top pressure for the gray cloud contribution. With \texttt{pRT}, we use the condensation profiles to infer the base of the cloud deck of the considered cloud species (MgSiO$_3$ and Fe). Figure~\ref{fig:retrieval_pTs} shows that the cloud base is expected to be deeper in the atmosphere (around 20~bar) when considering the retrieved P-T profiles from \texttt{pRT}.

It is noteworthy that the C/O ratio is constrained with high precision (although a bimodal distribution is obtained with \texttt{pRT}), both with the grid and the free retrievals. This is surprising since the error bars on the C/O ratio are comparable with those obtained for $\beta$~Pic~b by \citet{gravity2020} at much better signal-to-noise in the K-band. For the grid retrievals with ER, the best-fit C/O ratios coincide with a discrete point in the model grid. This may indicate that the (linear) interpolation between grid points does not provide sufficiently accurate spectra such that the uncertainties on the C/O ratio from ER are expected to be underestimated. As mentioned above, the posterior distribution of the C/O ratio obtained with \texttt{pRT} is bimodal, which appears related to a correlation with the retrieved quenching pressure (cf. Figure~\ref{fig:posterior_pRT}). With $P_\mathrm{quench} \sim 1$~mbar, the retrieved C/O ratio is comparable to the value from the plain ER fit (i.e., $\sim 0.65$). For the second solution with $P_\mathrm{quench} \sim 10~\text{bar}$, the C/O ratio has a value of $\sim 0.9$ and is therefore consistent with \texttt{ATMO}. The retrievals with \texttt{pRT} and \texttt{ATMO} use chemical equilibrium abundances to determine the absorber abundances in the atmosphere. Both models use three free parameters for this, but with different choices on the parametrization: \texttt{pRT} fits for the atmospheric metallicity ([Fe/H]) and then varies the C/O ratio by changing the oxygen abundance. In addition, it retrieves a quenching pressure, above which the abundances of CH$_4$, H$_2$O, and CO were held constant. \texttt{ATMO} retrieves the atmospheric metallicity [Fe/H], and allows the C and O abundances to vary freely. The deep quenching pressure that was retrieved with \texttt{pRT} is located below the photosphere. Hence, the combination of $P_\mathrm{quench}$, [Fe/H], and C/O is used as a knob for changing the relative CO/CH$_4$/H$_2$O abundances in the photospheric region. In a similar way, the retrieval with \texttt{ATMO} allows for changing the C and O abundances directly, which appears to provide a similar freedom with the abundances retrieval, and, in the retrievals presented here, leads to a similar C/O ratio.

The comparison between the free retrievals presented here shows that a careful vetting of different input model assumptions is important. While retrieving a quenching pressure in the atmosphere appears physically justified (as applied by \texttt{pRT}), as does an independent variation in the oxygen and carbon content in the atmosphere (as applied by \texttt{ATMO}). The fact that both retrievals find consistent C/O constraints could be pure coincidence, and in principle combining both approaches may be best, but it is important to assess whether a retrieval model with a correspondingly increased number of free parameters is justified, for example by considering Bayes factors. Another important difference between the two retrieval models is the parametrization of both the atmospheric $P$--$T$ structure and the clouds, which are closely linked. The $P$-$T$ parametrization of \texttt{pRT} can in principle allow for more isothermal $P$-$T$ structures, while the parametrization of \texttt{ATMO}, which fixes the equilibrium temperature of the planet, will always lead to a non-negligible temperature gradient in the photosphere. As has been shown by, for example, \citet{tremblin2015,tremblin2016}, isothermal atmospheres may mimic clouds, and lead to cloud-free atmospheres even when (synthetic) cloudy spectra are fed into retrievals \citep{molliere2020}. This may explain why the $P$-$T$ profile retrieved by \texttt{pRT} is more isothermal than that retrieved by \texttt{ATMO}, but we note that both retrievals constrain the atmospheres to be very cloudy. Additionally, the comparative ease with which the \texttt{ATMO} parametrization can lead to cloudy spectra may be important because in \texttt{pRT} only specific combinations of the physically motivated cloud parameters (settling parameter $f_{\rm sed}$, atmospheric mixing strength $K_{zz}$, atmospheric $P$-$T$ profile in comparison to saturation vapor pressure curve of condensates) lead to cloudy solutions.

Apart from atmospheric clouds, another possible source of reddening could be an inclined circumplanetary disk. Evidence for circumplanetary accretion disks around young ($< 10~\text{Myr}$) giant planets has been seen in both planet formation simulations \citep[e.g.,][]{lubow1999,dangelo2002} and observations \citep[e.g.,][]{bowler2011,christiaens2019}. However, recent detections of long-lived accretion disks (``Peter-Pan disks'') around planetary-mass, brown dwarf, and M dwarf objects with ages of $\sim 20$--$55~\text{Myr}$ \citep{eriksson2020,boucher2016,murphy2018,lee2020,silverberg2020} suggest that a disk around HD~206893~B could be plausible if its age is at the lower end of the range of $\sim 50$--$700~\text{Myr}$ predicted by \citet{delorme2017}. The interferometric visibilities measured by GRAVITY agree with the SPHERE K1- and K2-band photometry within 11.5\% and 7.7\%, respectively, and we find no dependence of the visibility as a function of the baseline length. Assuming emission from a uniform disk, we thus find an upper limit on the disk diameter of $\sim 89~\text{R}_\text{Jup}$ and $\sim 77~\text{R}_\text{Jup}$, respectively, with a longest baseline of $130~\text{m}$ \citep[cf. also][who did a similar exercise for the PDS~70 protoplanets]{wang2021}.

\subsection{Mass}
\label{sec:mass}

From the comparison between the best fit atmospheric parameters and the AMES-Cond evolutionary tracks in Section~\ref{sec:evolutionary_tracks}, we find that the dusty ER grid would be consistent with an extremely young object and the plain DP grid would be consistent with a rather old object. In between the two, the dusty BT grid suggests a moderately young ($\sim 3$--$300~\text{Myr}$) object somewhere between $\sim 5$--$30~\text{M}_\text{Jup}$. While the uncertainties on HD~206893~B's age and mass are large, it is still noteworthy that this solution is also in agreement with both the age of $\sim 40$--$270~\text{Myr}$ estimated for the Argus moving group \citep{torres2008,bell2015} and the mass of $10^{+5}_{-4}~\text{M}_\text{Jup}$ predicted from radial velocity and \emph{Gaia} data of the system \citep{grandjean2019}. Compared to \citet{delorme2017}, who report a best fit age and mass of $100$--$300~\text{Myr}$ and $15$--$30~\text{M}_\text{Jup}$, respectively, we find that a younger, planetary-mass object would fit the data, too, although this hypothesis hinges on a moderately low Argus moving group membership probability and the dynamical mass estimate from radial velocity and \emph{Gaia} data is likely biased due to the presence of the inner companion HD~206893~C. A low surface gravity is supported by the spectral retrievals with \texttt{petitRADTRANS} and \texttt{ATMO}, whose best fit atmospheric parameters are roughly consistent with those predicted by the dusty ER grid, but \citet{allers2013} mention that older and dusty field brown dwarfs can have very similar H- and K-band spectral features than young low-gravity objects.

By analyzing the \emph{Gaia} proper motion anomaly of the system (cf. Section~\ref{sec:gaia_proper_motion_anomaly}), we obtain an independent constraint on the mass of HD~206893~B and find that a second, closer-in companion (HD~206893~C) is required to explain the observed data in accordance with \citet{grandjean2019}. The mass range predicted for HD~206893~B is consistent with the $5$--$30~\text{M}_\text{Jup}$ obtained from the comparison between the best fit atmospheric parameters and evolutionary tracks, except for the case where the two companions are located roughly on the opposite side of their orbits. In this case, their masses remain poorly constrained. The mass of HD~206893~C should be between $\sim 8$--$15~\text{M}_\text{Jup}$ and depending on how well the mass of HD~206893~B is known, the on-sky position of HD~206893~C could be narrowed down to a parameter space where searching for it with the GRAVITY instrument could be feasible.


\section{Conclusions}
\label{sec:conclusions}

We present new VLTI/GRAVITY K-band spectroscopy at a resolution of $R \sim 500$ of the reddest known substellar companion HD~206893~B. From these observations we obtain two new astrometric data points with a precision of $\sim 100~\text{\textmu as}$ as well as a low signal-to-noise K-band spectrum of HD~206893~B. We use the astrometry to update the orbital parameters of HD~206893~B and the spectrum to infer its atmospheric parameters with atmospheric model fits and spectral retrievals. Given the previously observed difficulties with fitting both the extremely red color as well as the very shallow $1.4~\text{\textmu m}$ water absorption feature of HD~206893~B \citep{delorme2017,ward-duong2020}, we include additional extinction by high-altitude dust clouds made of enstatite grains in our atmospheric model fits.

From the orbit fits, we find that the GRAVITY data resolves the previously observed degeneracy between a lower eccentricity, larger semimajor axis and a higher eccentricity, smaller semimajor axis orbit by preferring the latter of these. The orbital solution for HD~206893~B preferred by the GRAVITY data suggests a mutual inclination of $20.8^{+13.6}_{-11.2}~\text{deg}$ between the orbit of HD~206893~B and the debris disk of the system. We argue that a misalignment between them could suggest a significantly younger age for the system or could be caused by dynamical planet-planet interactions with other putative companions in the system. However, such a misalignment needs to be confirmed by future GRAVITY observations and a profound analysis of dynamical planet-planet interactions and their impact on the alignment and eccentricity of HD~206893~B's orbit is left for future work.

From the grid retrievals, we find that the BT-Settl-CIFIST (BT) and Exo-REM (ER) grids including additional extinction (dusty models) can fit all near-infrared spectral features of HD~206893~B, namely the extremely red color, the very shallow $1.4~\text{\textmu m}$ water absorption feature, and the pointy H-band peak observed with GPI. However, both DRIFT-PHOENIX (DP) grids with (dusty models) and without (plain models) additional extinction fail to reproduce the pointy H-band peak. By comparison to evolutionary tracks, we argue that only the best fit dusty BT and ER models as well as the plain DP model correspond to physically plausible objects. The best fit parameters of these three models are spread over a wide range of ages and masses, though. If the best fit dusty BT and ER models are favored over the best fit plain DP model, due to their ability to fit the pointy H-band peak observed with GPI, we predict an age of $\sim 3$--$300~\text{Myr}$ and a mass of $\sim 5$--$30~\text{M}_\text{Jup}$ for HD~206893~B. This mass estimate from atmospheric models and evolutionary tracks is consistent with the mass estimate from the \emph{Gaia} proper motion anomaly of the system.

From the free retrievals with \texttt{petitRADTRANS} (\texttt{pRT}) and \texttt{ATMO}, we obtain parameters that are roughly consistent with those predicted by the dusty BT and ER models. Most notably are a very low surface gravity, high metallicity, and high C/O ratio. While the atmospheric chemistry and formation of clouds are handled differently between \texttt{pRT} and \texttt{ATMO}, we find consistent C/O ratios of $\sim 0.8$--$0.9$. However, their high precision is surprising given the much worse signal-to-noise in the K-band if compared to $\beta$~Pic~b \citep{gravity2020}, for which a similar precision was achieved. This might hint at the C/O constraints for HD~206893~B being driven by systematic effects.

The low surface gravity together with the mass estimate of $10^{+5}_{-4}~\text{M}_\text{Jup}$ from radial velocity and \emph{Gaia} data \citep{grandjean2019} and a potential Argus moving group membership of the system \citep[membership probability $\sim 61\%$,][]{ward-duong2020} suggests that a planetary nature could be possible for HD~206893~B. While tension exists across the various age indicators, with the extensive host star analysis performed by \citet{delorme2017} showing both young (rotational period), old (lithium and barium abundance) and ambiguous (chromospheric activity) indicators, the most recent available kinematic analyses from Banyan Sigma and \citet{lee2019} point to a younger age and potential membership with Argus \citep{zuckerman2019}. In addition, disentangling youth and low gravity from a dusty atmosphere can be difficult in the L-dwarf regime \citep{allers2013}, so that further observations such as more precise L and M-band photometry from the \emph{James Webb Space Telescope} and a broader spectral coverage or higher spectral resolution are required to make a robust statement on the nature of HD~206893~B. In agreement with \citet{grandjean2019}, we also find that the \emph{Gaia} proper motion anomaly of the system suggests a second, closer-in companion (HD~206893~C) whose mass could be in the planetary-mass regime.

Finally, it has been shown that the extreme atmospheric conditions on HD~206893~B responsible for its exceptionally red color cannot be reproduced by currently available atmospheric models for giant planets and brown dwarfs without further adaptions. The case of HD~206893~B can therefore serve as a benchmark for the further development of such atmospheric models which could ultimately lead to a more complete understanding of the objects at the boundary between exoplanets and brown dwarfs. Moreover, future radial velocity or high-contrast imaging observations might confirm the additional companions predicted to exist in this system and improve the mass estimate for HD~206893~B.

\begin{acknowledgements}

The authors would like to thank Rob~De~Rosa for his insights on orbit fitting and the comparison with the GPI results. This research has made use of the Jean-Marie Mariotti Center \texttt{Aspro} service \footnote{http://www.jmmc.fr/aspro}. P.~M. and T.~H. acknowledge support from the European Research Council under the Horizon 2020 Framework Program via the ERC Advanced Grant Origins 83 24 28. T.~S. acknowledges support from the Netherlands Organisation for Scientific Research (NWO) through grant VI.Veni.202.230. This work was performed using the ALICE compute resources provided by Leiden University. A.~V., G.~O., and M.~H. acknowledge funding from the European Research Council (ERC) under the European Union's Horizon 2020 research and innovation program (grant agreement no.~757561). The manuscript was also substantially improved following helpful comments from an anonymous referee.

\end{acknowledgements}

\bibliographystyle{aa}
\bibliography{references}

\begin{appendix}

\section{Orbit fitting and debris disk}
\label{sec:orbit_fitting_and_debris_disk}

\begin{table*}
\caption{Relative astrometry of HD~206893~B from the literature. ``Sep.'' and ``PA'' denote the angular separation and the position angle, respectively, and ``Inst.'' denotes the instrument with which the data were acquired.}
\label{tab:hd206893b_astrometry}
\centering
\begin{tabular}{cccccc}
\hline\hline
MJD & Sep. & PA & Inst. & Band & Source \\
(days) & (mas) & (deg) & & & \\
\hline
57300 & $270.0 \pm 2.6$ & $69.95 \pm 0.55$ & SPHERE & H & M17 \\
57608 & $269.53 \pm 12.15$ & $62.76 \pm 2.16$ & NACO & L' & S20 \\
57647 & $265 \pm 2$ & $62.25 \pm 0.11$ & SPHERE & K1/K2 & D17 \\
57653 & $267.6 \pm 2.9$ & $62.72 \pm 0.62$ & GPI & H & W21 \\
57682 & $265.0 \pm 2.7$ & $61.33 \pm 0.64$ & GPI & K1 & W21 \\
57948 & $260.3 \pm 2.0$ & $54.2 \pm 0.4$ & SPHERE & H & G19 \\
58066 & $256.9 \pm 1.1$ & $51.01 \pm 0.35$ & GPI & K2 & W21 \\
58277 & $246.51 \pm 21.34$ & $42.80 \pm 2.24$ & NACO & NB4.05 & S20 \\
58289 & $249.11 \pm 1.60$ & $45.50 \pm 0.37$ & SPHERE & H2/H3 & G19 \\
58385 & $251.7 \pm 5.4$ & $42.6 \pm 1.6$ & GPI & J & W21 \\
58415 & $239.12 \pm 17.55$ & $42.53 \pm 2.17$ & NACO & M' & S20 \\
\hline
\multicolumn{6}{l}{\textbf{Notes.} \parbox[t]{9 cm}{M17 = \citet{milli2017}, D17 = \citet{delorme2017}, G19 = \citet{grandjean2019}, S20 = \citet{stolker2020}, W21 = \citet{ward-duong2020}.}}
\end{tabular}
\end{table*}

\begin{figure*}
\centering
\includegraphics[width=0.85\textwidth]{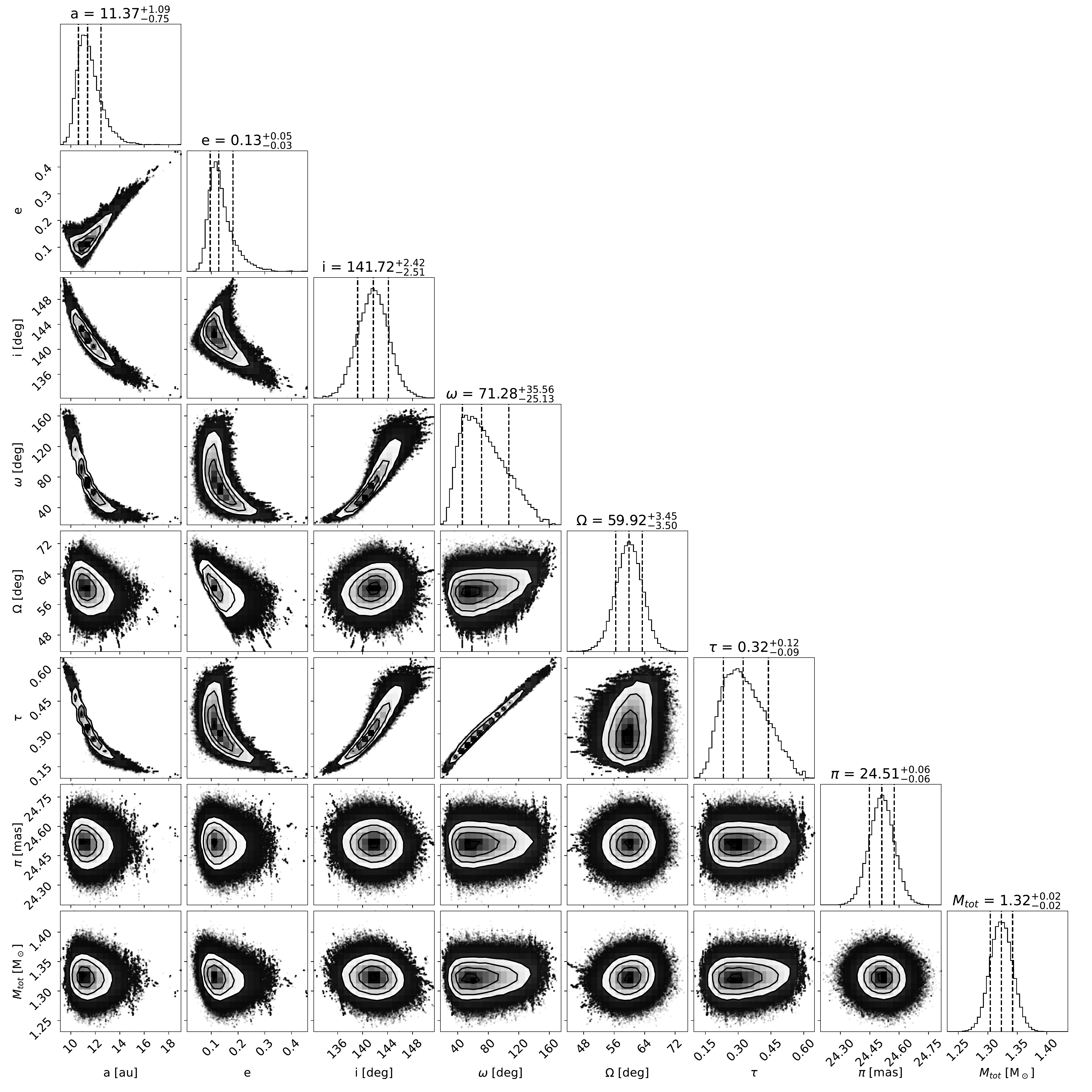}
\includegraphics[width=0.85\textwidth]{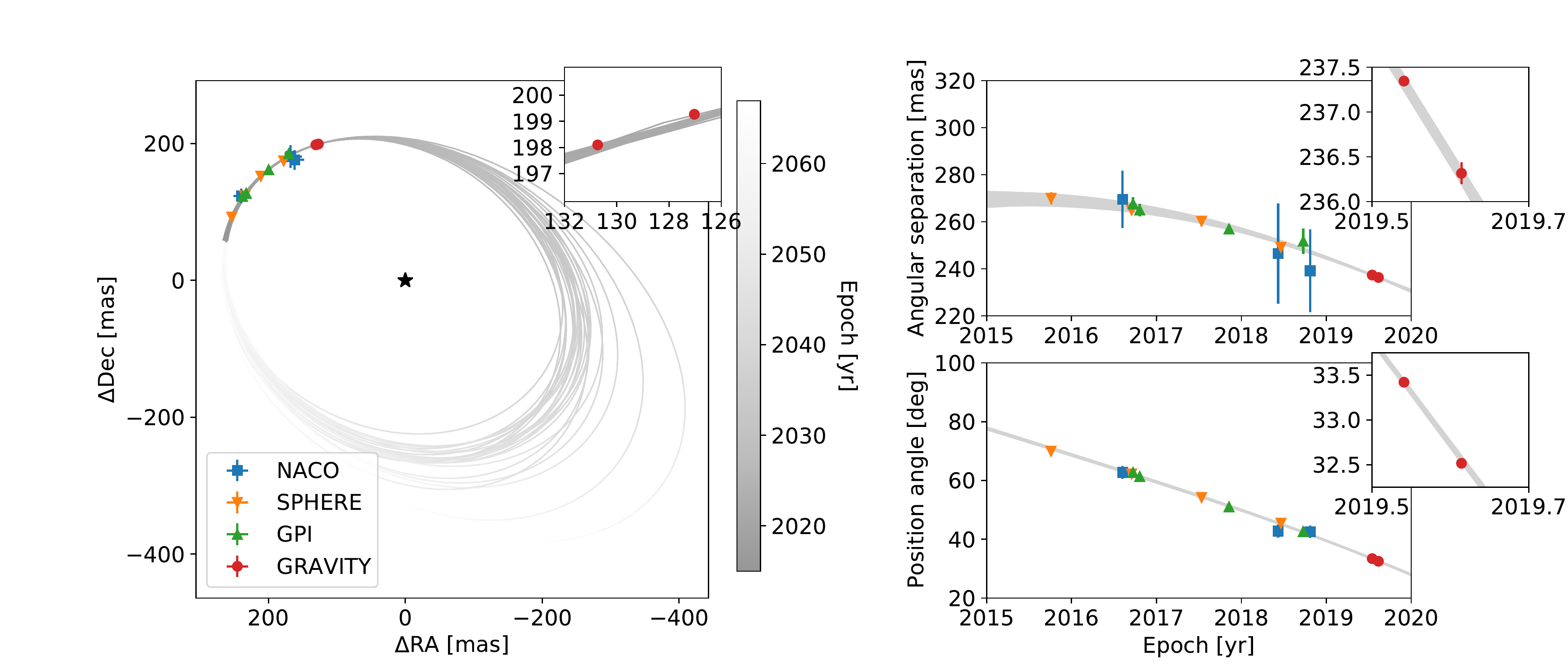}
\caption{Same as Figure~\ref{fig:orbit_fitting_free}, but using Gaussian priors of $140 \pm 3~\text{deg}$ for the inclination $i$ and $61 \pm 4~\text{deg}$ for the longitude of the ascending node $\Omega$ in order to enforce coplanarity between the orbit of HD~206893~B and the debris disk of the system.}
\label{fig:orbit_fitting_disk}
\end{figure*}

\section{Photometry}
\label{sec:photometry}

\begin{table}[H]
\caption{Photometry of HD~206893~B from the literature. ``Inst.'' denotes the instrument with which the data were acquired.}
\label{tab:hd206893b_photometry}
\centering
\begin{tabular}{ccccc}
\hline\hline
Inst. & Band & $\lambda$ & Value & Source \\
& & ($\text{\textmu m}$) & (mag) & \\
\hline
SPHERE & J & 1.245 & $18.33 \pm 0.17$ & D17 \\
SPHERE & H & 1.626 & $16.79 \pm 0.06$ & D17 \\
SPHERE & K1 & 2.104 & $15.20 \pm 0.10$ & D17 \\
SPHERE & K2 & 2.255 & $14.88 \pm 0.09$ & D17 \\
NACO & L' & 3.805 & $13.80 \pm 0.31$ & S20 \\
NACO & NB405 & 4.056 & $13.17 \pm 0.55$ & S20 \\
NACO & M' & 4.781 & $12.78 \pm 0.51$ & S20 \\
\hline
\multicolumn{5}{l}{\textbf{Notes.} \parbox[t]{6.5 cm}{D17 = \citet{delorme2017}, S20 = \citet{stolker2020}.}}
\end{tabular}
\end{table}

\section{Atmospheric model fitting posteriors}
\label{sec:atmospheric_model_fitting_posteriors}

\begin{figure*}
\centering
\includegraphics[width=\textwidth]{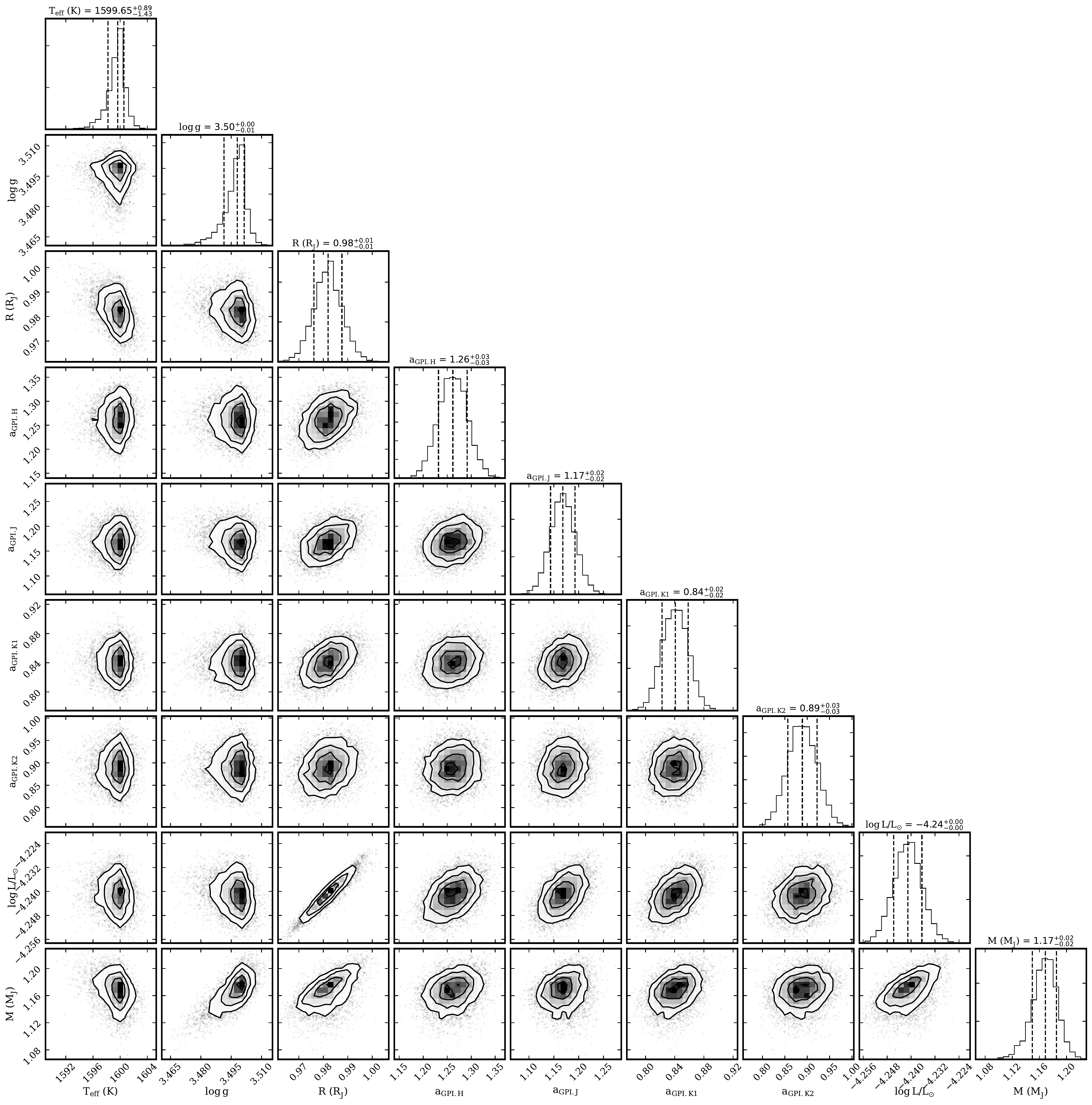}
\caption{Posterior distribution of the atmospheric parameters of the plain BT-Settl-CIFIST model fitted to the SPHERE, GPI, and GRAVITY spectra and the NACO and SPHERE photometry. The values state the 68\% confidence intervals around the median.}
\label{fig:posterior_BT}
\end{figure*}

\begin{figure*}
\centering
\includegraphics[width=\textwidth]{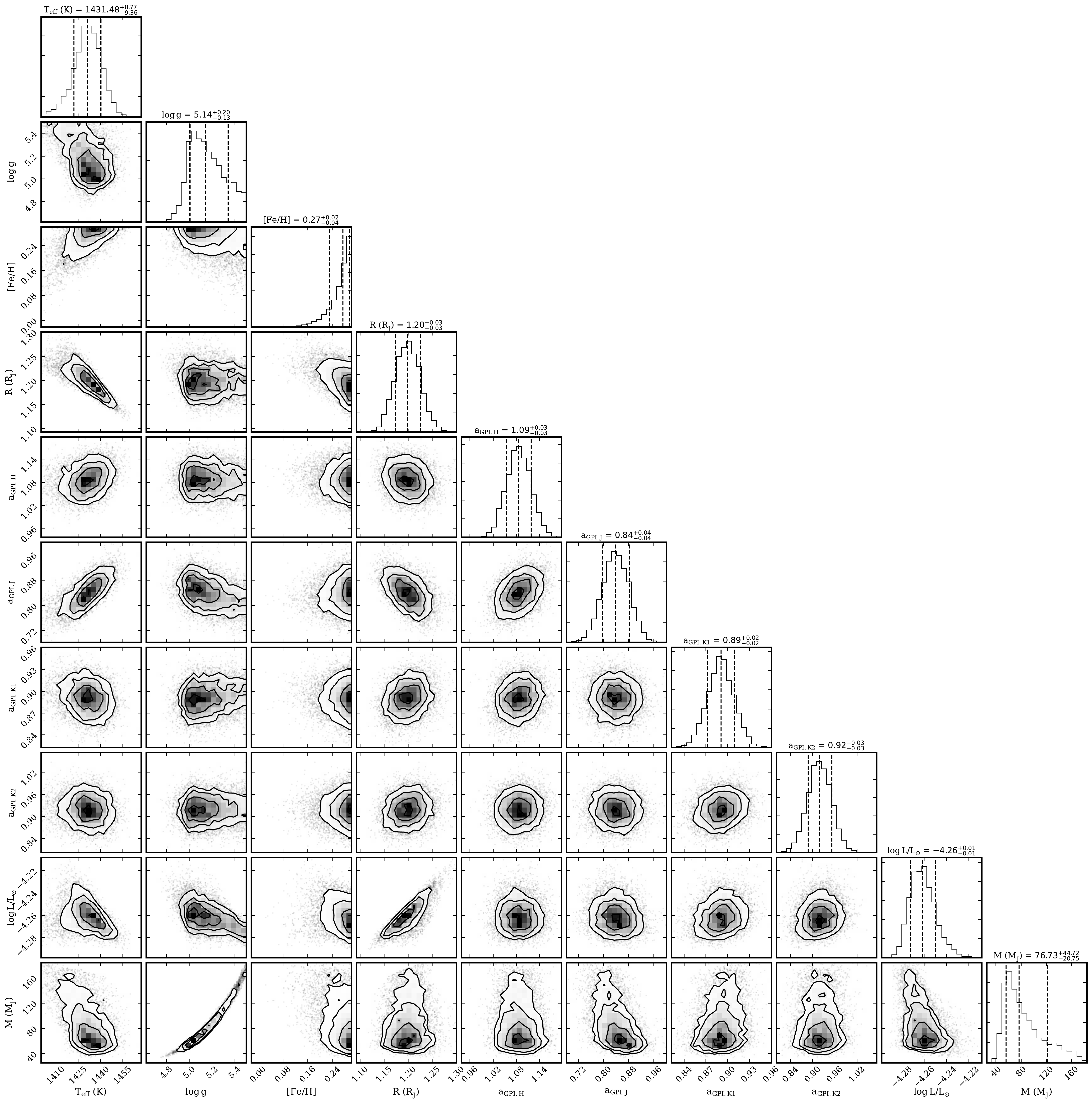}
\caption{Same as Figure~\ref{fig:posterior_BT}, but for the plain DRIFT-PHOENIX model.}
\label{fig:posterior_DP}
\end{figure*}

\begin{figure*}
\centering
\includegraphics[width=\textwidth]{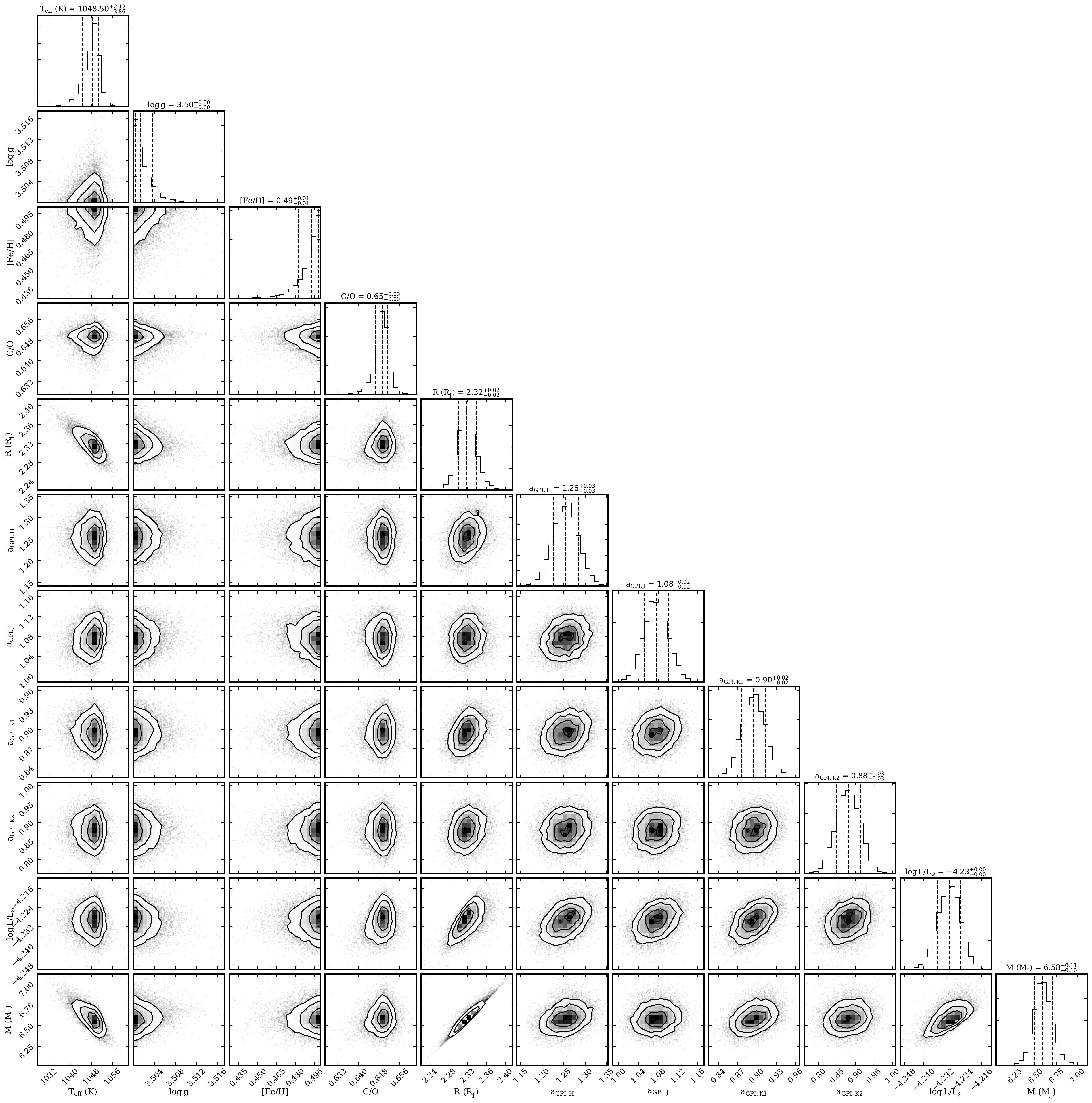}
\caption{Same as Figure~\ref{fig:posterior_BT}, but for the plain Exo-REM model.}
\label{fig:posterior_ER}
\end{figure*}

\begin{figure*}
\centering
\includegraphics[width=\textwidth]{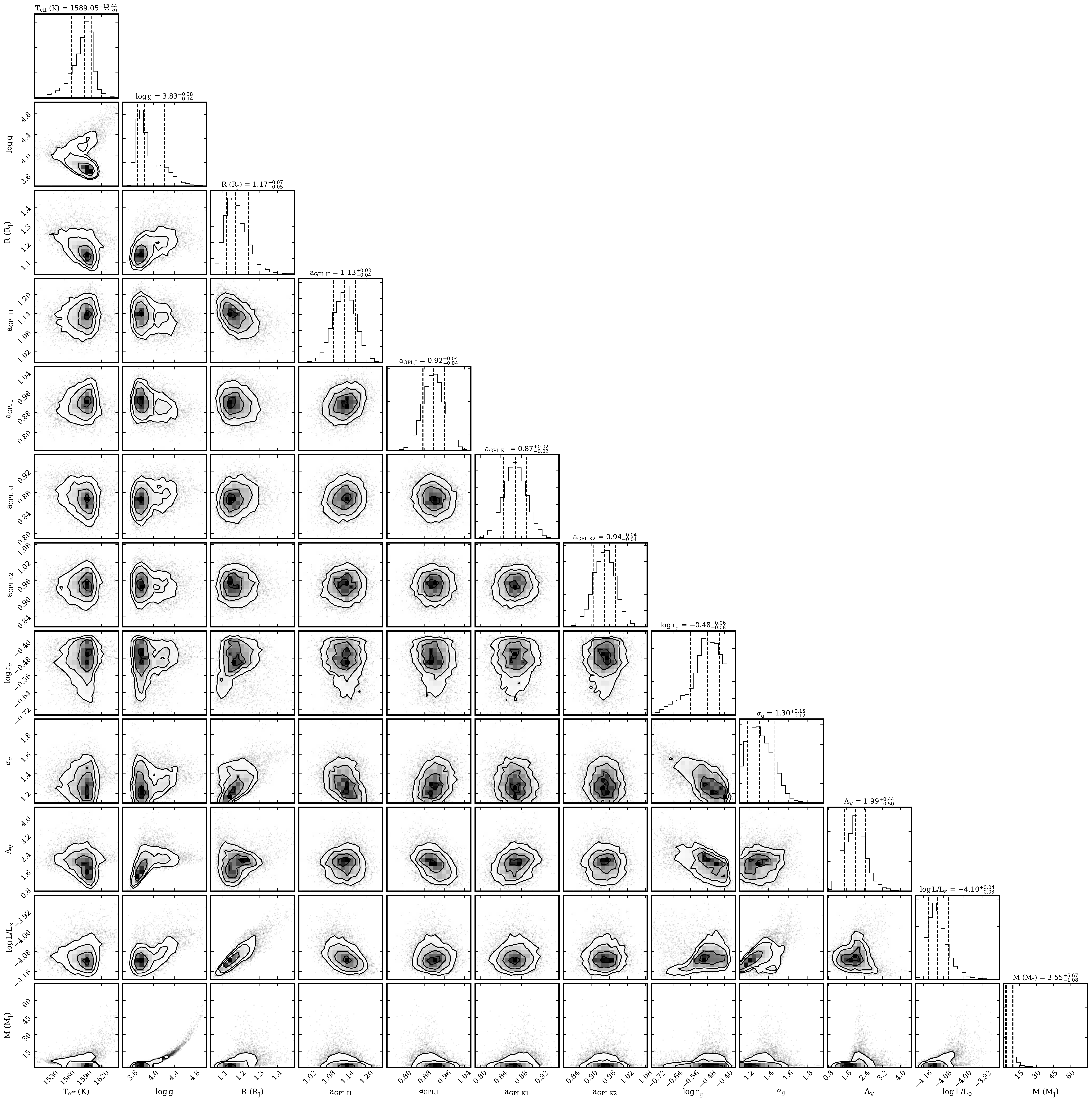}
\caption{Same as Figure~\ref{fig:posterior_BT}, but for the dusty BT-Settl-CIFIST model. We note that the luminosity is calculated from the effective temperature and the radius and is no longer consistent with the luminosity that would be calculated from the reddened spectrum (i.e., the calculated luminosity is larger).}
\label{fig:posterior_BT_dust}
\end{figure*}

\begin{figure*}
\centering
\includegraphics[width=\textwidth]{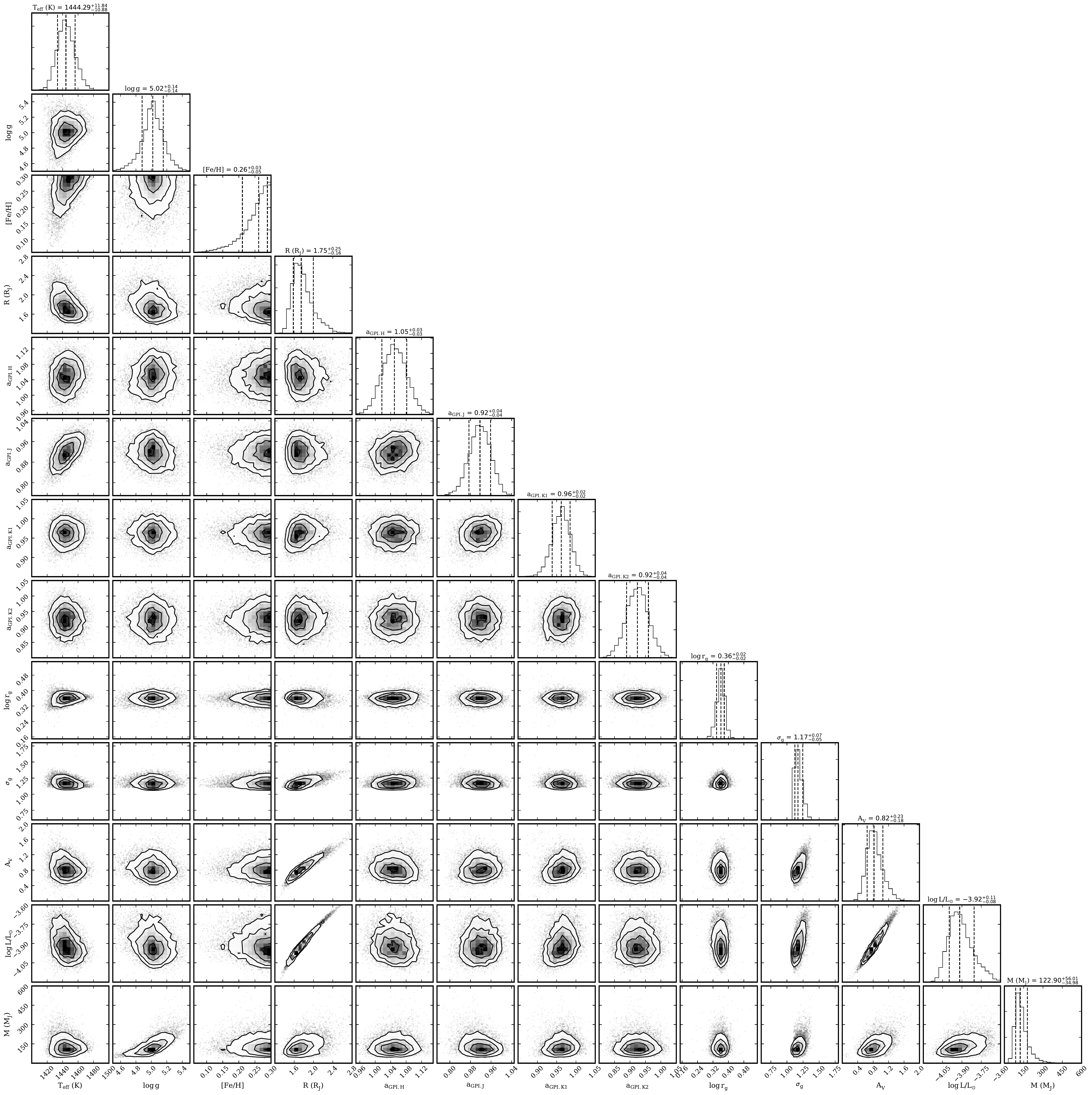}
\caption{Same as Figure~\ref{fig:posterior_BT_dust}, but for the dusty DRIFT-PHOENIX model.}
\label{fig:posterior_DP_dust}
\end{figure*}

\begin{figure*}
\centering
\includegraphics[width=\textwidth]{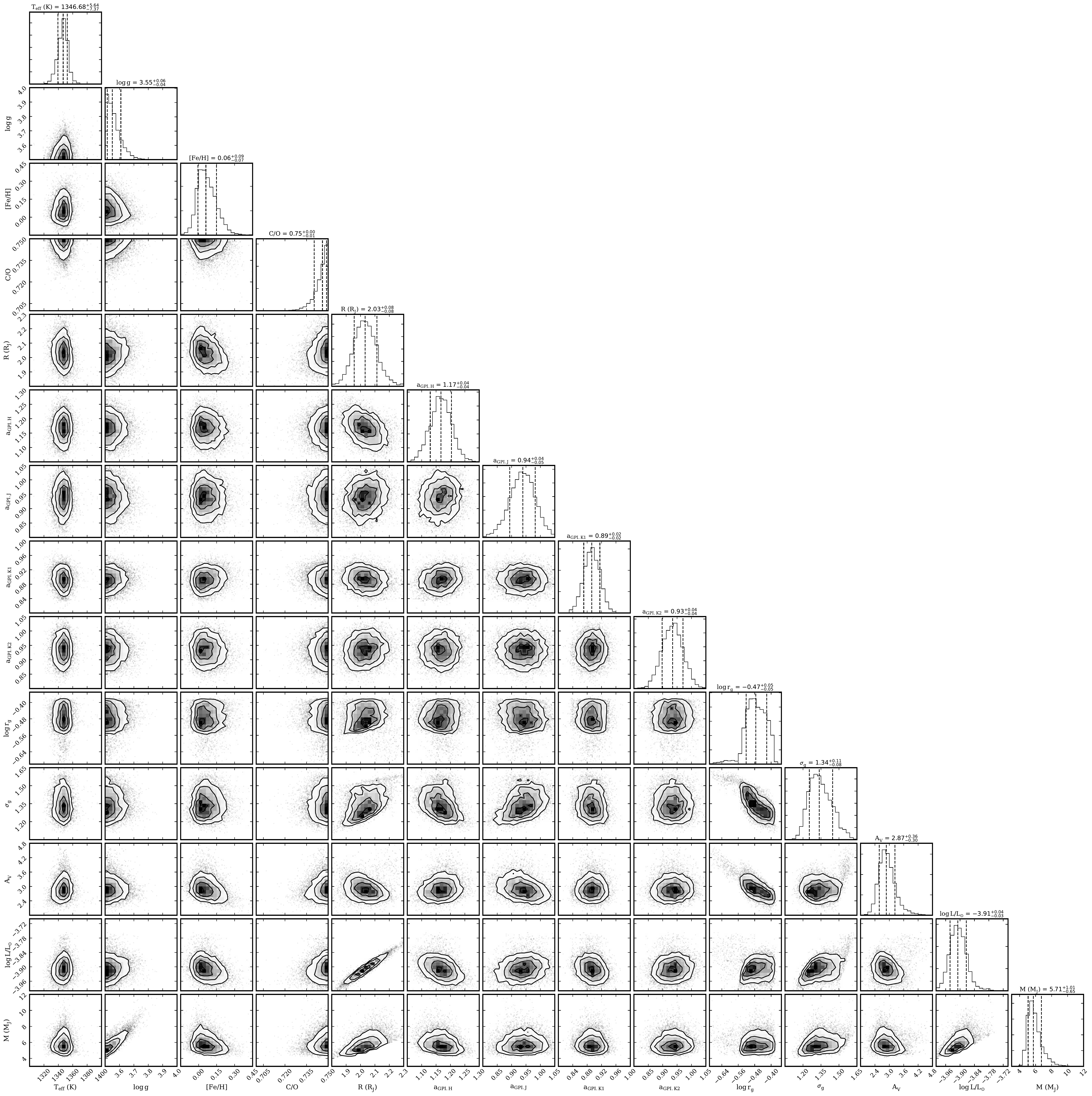}
\caption{Same as Figure~\ref{fig:posterior_BT_dust}, but for the dusty Exo-REM model.}
\label{fig:posterior_ER_dust}
\end{figure*}

\begin{figure*}
\centering
\includegraphics[width=\textwidth]{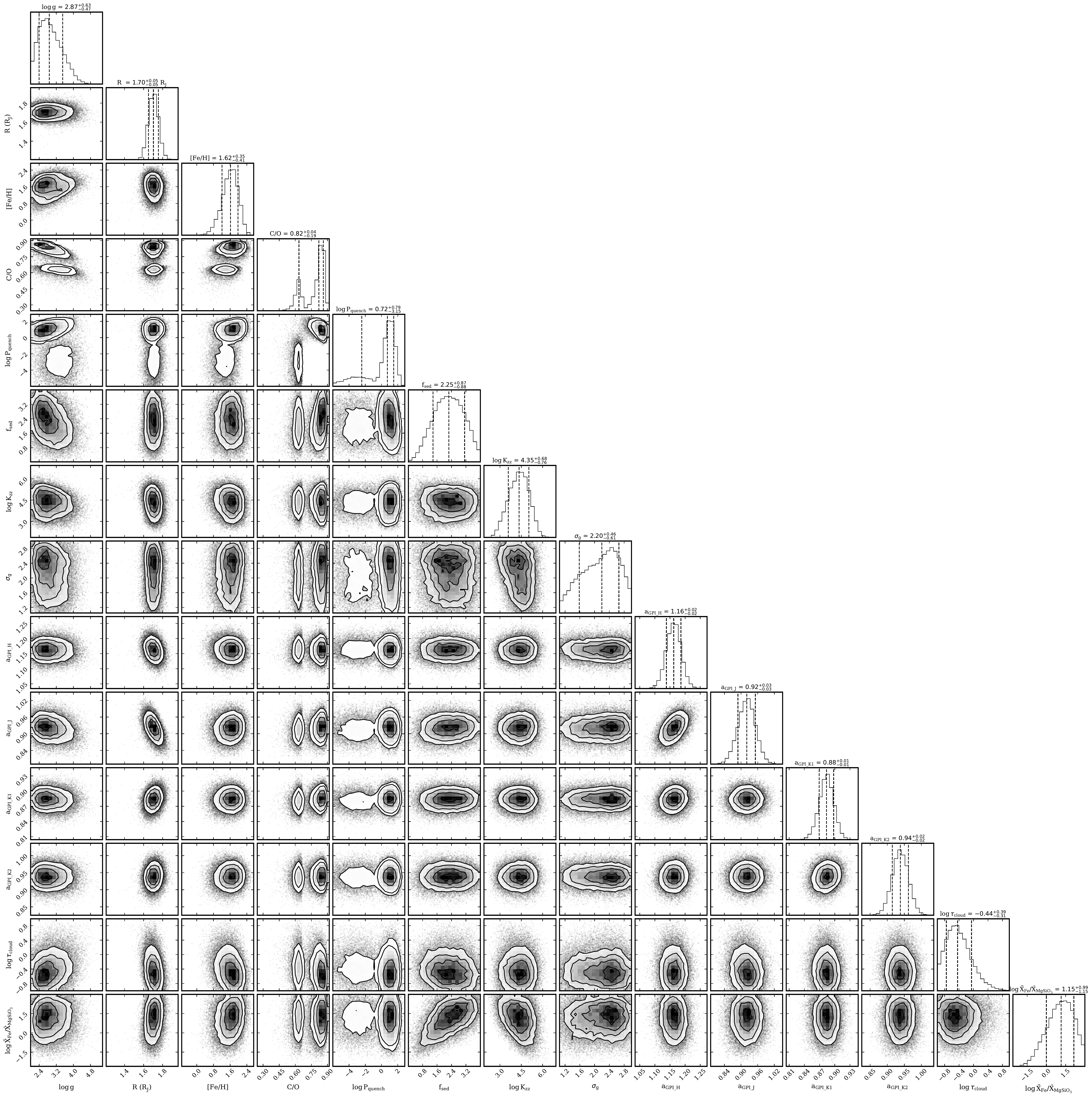}
\caption{Posterior distribution of the spectral retrieval with \texttt{petitRADTRANS}.}
\label{fig:posterior_pRT}
\end{figure*}

\begin{figure*}
\centering
\includegraphics[width=\textwidth]{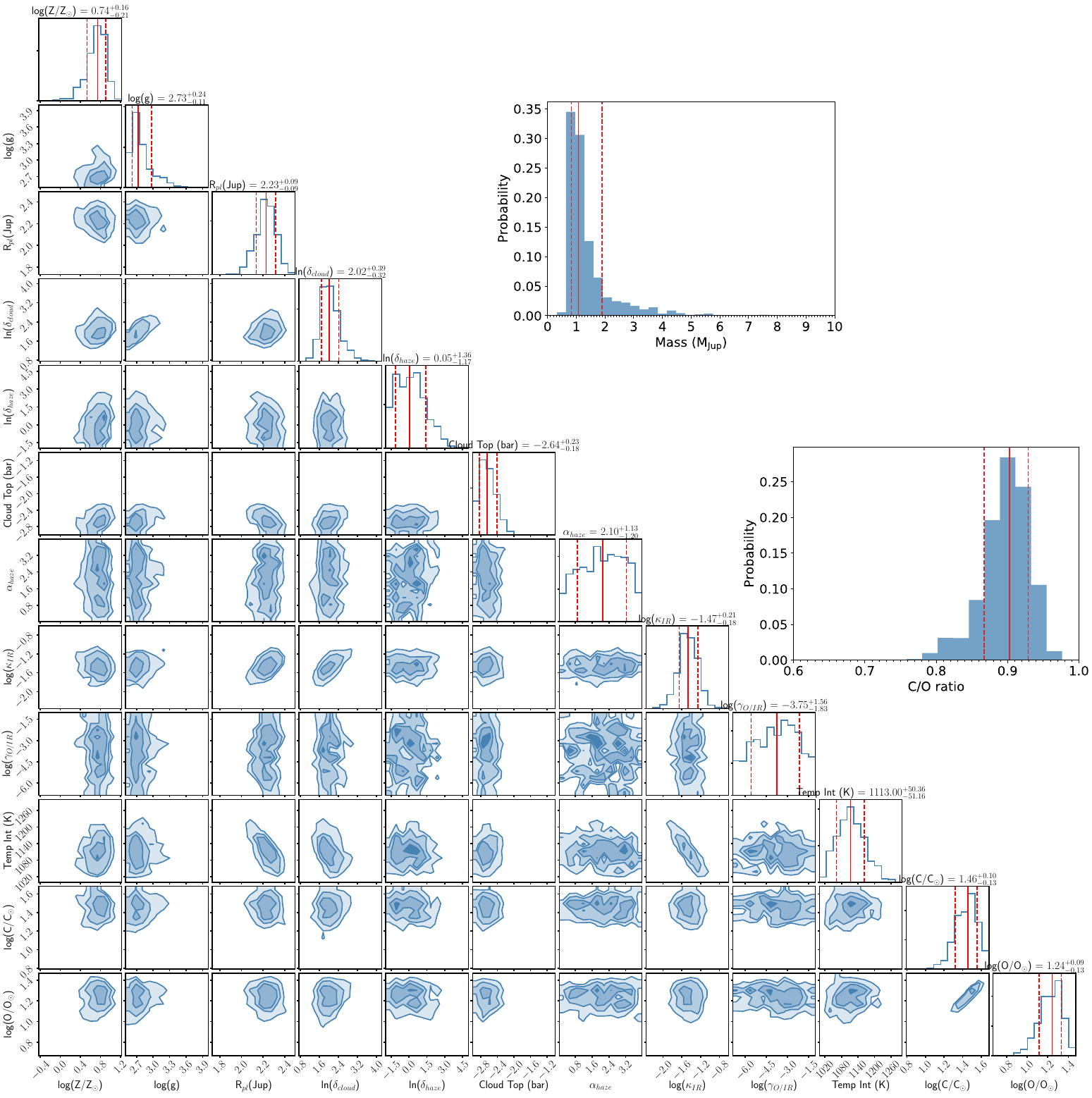}
\caption{Posterior distribution of the spectral retrieval with \texttt{ATMO}.}
\label{fig:posterior_ATMO}
\end{figure*}

\end{appendix}

\end{document}